\documentstyle[12pt,fleqn]{article}
\renewcommand{\theequation}{\thesection.\arabic{equation}}
\renewcommand{\thefootnote}{\fnsymbol{footnote}}
\newlength{\extraspace}
\newlength{\extraspaces}
\newcommand{\be}{\begin{equation}
\addtolength{\abovedisplayskip}{\extraspaces}
\addtolength{\belowdisplayskip}{\extraspaces}
\addtolength{\abovedisplayshortskip}{\extraspace}
\addtolength{\belowdisplayshortskip}{\extraspace}}
\newcommand{\ee}{\end{equation}}

\newcommand{\ba}{\begin{eqnarray}
\addtolength{\abovedisplayskip}{\extraspaces}
\addtolength{\belowdisplayskip}{\extraspaces}
\addtolength{\abovedisplayshortskip}{\extraspace}
\addtolength{\belowdisplayshortskip}{\extraspace}}
\newcommand{\ea}{\end{eqnarray}}

\newcommand{\bas}{\begin{eqnarray*}
\addtolength{\abovedisplayskip}{\extraspaces}
\addtolength{\belowdisplayskip}{\extraspaces}
\addtolength{\abovedisplayshortskip}{\extraspace}
\addtolength{\belowdisplayshortskip}{\extraspace}}
\newcommand{\eas}{\end{eqnarray*}}

\newenvironment{theorem}[1]
{ \vspace{3mm}\noindent {\bf #1 :} }{\vspace{2mm}}
\newcommand{\bt}[1]{\begin{theorem}{#1}}
\newcommand{\et}{\end{theorem}}

\newcommand{\newsection}[1]{
\vspace{15mm}
\pagebreak[3]
\addtocounter{section}{1}
\setcounter{equation}{0}
\setcounter{subsection}{0}
\setcounter{footnote}{0}
\begin{flushleft}
{\large\bf \thesection. #1}
\end{flushleft}
\nopagebreak
\medskip
\nopagebreak}

\newcommand{\newsubsection}[1]{
\vspace{1cm}
\pagebreak[3]
\addtocounter{subsection}{1}
\begin{flushleft}
{ \bf \thesubsection. #1}
\end{flushleft}
\nopagebreak
\smallskip
\vspace{-4mm}
\nopagebreak}

\newcommand{\newappendix}[1]{
\vspace{15mm}
\pagebreak[3]
\addtocounter{section}{1}
\setcounter{equation}{0}
\setcounter{subsection}{0}
\setcounter{footnote}{0}
\renewcommand{\theequation}{\Alph{section}.\arabic{equation}}
\begin{flushleft}
{\large\bf \Alph{section}. #1}
\end{flushleft}
\nopagebreak
\medskip
\nopagebreak}

\newcounter{prop}

\newcommand{\NP}[1]{Nucl.\ Phys.\ {\bf #1}}
\newcommand{\PL}[1]{Phys.\ Lett.\ {\bf #1}}
\newcommand{\CMP}[1]{Comm.\ Math.\ Phys.\ {\bf #1}}

\newcommand{\N}{\mbox{I\hspace{-.4ex}N}}
\newcommand{\C}{\,\mbox{{\sf I}\hspace{-1.2ex}{\bf C}}}
\newcommand{\Z}{\mbox{Z\hspace{-.9ex}Z}}
\newcommand{\R}{\mbox{\rm I\hspace{-.4ex}R}}
\newcommand{\bra}{\langle\,}
\newcommand{\ket}{\,\rangle}
\newcommand{\ra}{\rightarrow}

\newcommand{\rra}{\ \longrightarrow \ }

\newcommand{\is}{\! & \! = \! & \!}
\newcommand{\nonum}{\nonumber \\[1.5mm]}
\newcommand{\sspace}{\makebox[1cm]{ }}
\newcommand{\bspace}{\makebox[2cm]{ }}
\newcommand{\nspace}{\!\!\!\!\!\!\!\!\!\!}

\newcommand{\Tr}{{\rm Tr}}

\newcommand{\lb}{\lambda}
\renewcommand{\d}{{\partial}}
\newcommand{\dd}[1]{ \frac{\partial}{\partial{#1}} }
\newcommand{\ddx}[1]{ \frac{\partial}{\partial x_{#1}} }
\newcommand{\dpl}{\partial_{+}}
\newcommand{\dmi}{\partial_{-}}

\newcommand{\fh}{{\widehat{f}}}
\newcommand{\gh}{{\widehat{g}}}
\newcommand{\mh}{{\widehat{m}}}

\newcommand{\Gh}{{\widehat{G}}}

\newcommand{\cF}{{\cal F}}

\newcommand{\cL}{{\cal L}}

\newcommand{\cO}{{\cal O}}

\newcommand{\Inmi}{I^{(n)}_-}
\newcommand{\Inpl}{I^{(n)}_+}
\newcommand{\In}{I^{(n)}}
\newcommand{\Jn}{J^{(n)}}
\newcommand{\Jm}{J^{(m)}}
\newcommand{\Gn}{G^{(n)}}
\newcommand{\gn}{g^{(n)}}
\newcommand{\Jnmi}{J^{(n)}_-}
\newcommand{\Jnpl}{J^{(n)}_+}
\newcommand{\ZZ}[2]{Z_{#1}(\theta_{#2})}
\newcommand{\Zt}[1]{Z_{#1}(\theta)}
\newcommand{\dmiphi}{\partial_-\Phi}
\newcommand{\sh}{\mbox{sh}}
\newcommand{\dis}{\displaystyle}
\newcounter{abc}
\begin{document}
%
\begin{titlepage}
%
\renewcommand{\thefootnote}{\fnsymbol{footnote}}
\begin{flushright}
MPI-Ph/93-92\\
November 1993
\end{flushright}
\vspace{1cm}

\begin{center}
{\LARGE The Quantum Spectrum of the\\
       $\;\,$Conserved Charges\\[3mm]
       $\;\;\;\;\;$in Affine Toda Theories}
{\makebox[1cm]{ }
          \\[2cm]
{\large Max R. Niedermaier}\\ [3mm]
{\small\sl Max-Planck-%
Institut f\"{u}r Physik} \\
{-\small\sl Werner Heisenberg Institut - } \\
{\small\sl F\"{o}hringer Ring 6}\\
{\small\sl 80805 Munich (Fed. Rep. Germany)}}
\vspace{3.5cm}

{\bf Abstract}
\end{center}
\begin{quote}
The exact eigenvalues of the infinite set of conserved
charges on the multi-particle states in affine Toda
theories are determined. This is done by constructing
a free field realization of the Zamolodchikov-Faddeev
algebra in which the conserved charges are realized as
derivative operators. The resulting eigenvalues are
renormalization group (RG) invariant, have the correct
classical limit and pass checks in first order
perturbation theory. For $n=1$ one recovers the
(RG invariant form of the) quantum masses of Destri and
DeVega.
\end{quote}
\vfill
\renewcommand{\thefootnote}{\arabic{footnote}}
\setcounter{footnote}{0}
\end{titlepage}
%
%
\newsection{Introduction}
The infinite set of local conserved charges in involution
is a key structure in an integrable relativistic
field theory. The simultaneous eigenstates are the
multiparticle states of the theory and the knowledge
of their spectrum is a major step towards the
linearization of the dynamics. In the classical case
the linearized dynamics can in principle be
constructed by means of the inverse scattering
transform. The transition to the quantum regime rests
on the construction of the renormalized monodromy
matrix satisfying a Yang Baxter equation. This however
has only been achieved in rare cases due to technical,
and perhaps fundamental obstructions to the applicability
of the method. This motivates the search for
alternative techniques.

The recent progress towards establishing such an
alternative has two major sources. The first is the
improved understanding of the form factor equations
in terms of infinite dimensional symmetry algebras
\cite{Yang,qKZ}. The second is a group of
ideas collectively
referred to as `perturbed conformal field theories'
\cite{Z,Trunc}. A central result there is that the local
conserved charges of an integrable massive QFT can
already be identified and constructed at the level of
the conformal field theory (CFT) describing its UV
behaviour \cite{Z,MN1,FF}.
The next step should consist in addressing the
diagonalization problem for these charges in the
context of perturbed CFTs and, in particular, to find
expressions for the eigenvalues. In the form factor
approach knowledge of the eigenvalues of the conserved
charges provides part of the additional dynamical input
required to set up the correspondence between solutions
of the form factor equations and local operators.
It is the purpose of the present paper to deduce exact
expressions for these eigenvalues on the multiparticle
states in affine Toda theories.

Affine Toda theories are relativistic models of $r$
interacting scalar fields which generalize the
Sinh/Sine Gordon model. They are associated with
the simple root system $\alpha_0,\ldots ,\alpha_r$
of some affine Lie algebra $\hat{g}$. The Lagrangian
in 2-dim Minkowski space is given by
\be
{\cal L}= \frac{1}{2}\partial_{\mu}\phi\cdot\partial^{\mu}
\phi - \frac{m^2}{\beta^2}\sum_{j=0}^r a_j
\left( e^{\beta \alpha_j\cdot\phi} -1 \right)\;,
\ee
where $a_j$ are the labels of the Dynkin diagram
(\cite{Kac}, p 54).
The fields $\phi^a(t,x),\; a=1,\;\ldots, r$  are
considered as the components of a real field with values
in the  Cartan subalgebra $h$ of $g$.
The coupling constant $\beta$ is real and $m$ fixes
the (bare) mass scale. At the classical level the theories
(1.1) are known to be integrable and an infinte set of
local conserved currents can be constructed by various
techniques \cite{OlTurok,Kyoto,DS}.
The application of the inverse scattering method
however meets certain obstructions. Nevertheless, the
principle aims of the method can still be achieved by
deriving a system of GLM-type equations whose solutions
are given in terms of generalized tau functions\cite{MN2}.
This leads to a parametrization and construction of
generic classical solutions in terms of `scattering data'
which linearize the dynamics. The classical spectral
problem then consists in the derivation of `trace
identities'. This means that one aims to find explicit
expressions for the local conserved charges in terms of the
scattering data i.e.
\bas
&& I^{(n)}[solution] = I^{(n)}(scattering\;data)\;,
\sspace n\in E\subset \N\;.
\eas
In section 4 we will recall such a trace identity \cite{MN2}
to check the classical limit of our quantum result.

In the quantum
theory, the natural analogue of a trace identity is the
sequence of eigenvalues of the conserved charges on the
asymptotic multi-particle states. In section 5 we will
use a vertex operator construction to obtain an exact
formula for these eigenvalues in the real coupling affine
Toda theories. Since the scattering operator commutes with
all the conserved charges $[S\,,\,\In]=0$, one expects
that the bootstrap $S$-matrix (once known) also carries
information about their spectrum. Indeed, the intertwining
concept is just that of a Zamolodchikov-Faddeev (ZF)
algebra. We call a ZF algebra $Z(S)$ associated with $S$ an
associative algebra with generators $\Zt{a},\;\theta\in \C,\;
\In,\;n\in E$ and $K$ subject to the relations
\ba
&& \ZZ{a}{a}\,\ZZ{b}{b} = S_{ab}(\theta_a -\theta_b)\,
\ZZ{b}{b}\,\ZZ{a}{a}\nonum
&& [\In\,,\,\Zt{a}]=e^{-n\theta}\,\In(a)\,\Zt{a}\;,
\bspace a,b=1,\ldots, r,\;\;n\in E\nonum
&& \left[ K\,,\; \Zt{a}\right] =
\frac{d}{d\theta}\,\Zt{a} \;.
\ea
$K$ is the generator of Lorentz boosts and $\In(a)$ is
interpreted as the eigenvalue of $\In$ on an asymptotic
1-particle state of type $a$.
Usually, a ZF algebra is supposed to act on the space of
scattering states of the theory $\Sigma_{in/out}$. These
are Fock spaces but the relation to the fundamental
fields of the theory is elusive in general. It is therefore
more useful to construct realizations
$\rho :Z(S)\rra \pi$
of (1.2) on some auxilary Fock space $\pi$, on which the
$\Zt{a}$'s act as generalized vertex operators. This,
of course, can be done in many ways and one will choose the
realization according to purpose. In section 5 we will show
that there exists a realization $\rho_I$ in which the
conserved charges are realized just as derivative operators
\be
\rho_I(\In) =\ddx{n}\;,\sspace n\in E\;.
\ee
Moreover this realization is shown to be essentially
uniquely determined by (1.3). Once the realization has
been constructed, the exact eigenvalues of the conserved
charges can be obtained from the second of the
relations (1.2). Since this is a novel technique,
we have included a number of independent checks
on the result. In particular, it turns out that
\begin{itemize}
\item[--] the eigenvalues obtained are
renormalization group (RG) invariant.
\item[--] they have the correct classical limit
and pass checks in 1st order perturbation theory.
\item[--] for $n=1$ one recovers the (RG invariant form
of the) exact quantum masses found by Destri and DeVega
\cite{DDeV}.
\end{itemize}
The paper is organized as follows. In section 3 we give
a quantum field theoretical definition of the eigenvalues
via a LSZ reduction formula, relating them to the Greens
functions of the conserved densities. From this one can
show that the eigenvalues are RG invariant and that
their functional forms (as a function of the coupling
constant $\beta$) are governed by  universal, scheme
independent functions $K_n(\beta)$. In section 4 we will
determine the classical limit of $K_n(\beta)$ and calculate
some first order quantum corrections. Finally, in section 5
the exact expression for $K_n(\beta)$ is obtained as
indicated. We conclude with some remarks on the
diagonalization problem of the conserved charges on the
Verma modules, which should be governed by `generalized
Kac determinants'.

\newsection{Real coupling affine Toda theories}
Here we prepare some basic results on real coupling
affine Toda theories (AT). Our kinematical conventions
are as follows: $(t,x)=(x^0,x^1)$ are coordinates on
2-dimensional Minkowski space $\R^{1,1}$ with metric
$\eta =\mbox{diag}(1,-1)$. Lightcone coordinates are
introduced by $x^{\pm}=(x^0\pm x^1)/\sqrt{2}$.
Elements $\Lambda$ of the restricted Lorentz group
$SO(1,1)$ are parametrized by the rapidity $\theta\in\R$
via $\Lambda(\theta)=e^{\theta K}$, where $K$ is the
generator of the Lie algebra. The $2^{n_++n_-}$ components
of a Lorentz tensor of type $(n_+,n_-)$ transform as
$(\Lambda t)_{+\ldots +,-\ldots -} = e^{n_+\theta}
e^{-n_-\theta}t_{+\ldots +,-\ldots -}$ in lightcone
coordinates. Indices are raised and lowered according to
$t^{+\cdots}=t_{-\cdots}$ and vice versa.
If $P_{\pm}$ are the lightcone
momenta, the Poincar\'{e} algebra takes the form
$[K,P_{\pm}]=\pm P_{\pm}$. On test functions one has the
usual realization in terms of differential operators
$iK= x^+\dpl -x^-\dmi,\;iP_{\pm} =\d_{\pm}$.

We will consider the case of simply laced Lie
algebras $g$, and mainly the $A$-series.
It is convenient to use a complex basis $T_1,\ldots, T_r$
for the Cartan subalgebra $h$, satisfying $(T_a,\,T_b)=
\delta_{a,\bar{b}},\; T^{\dagger}_a =T_{\bar{a}}$.
Here $(\;,\;)$ is the non-degenerate bilinear form
on $h$ and $\bar{a}$ is the charge conjugate of $a$.
The expansion coefficients $\Phi_a=
(T^{\dagger}_a,\,\Phi)$ are then complex fields
$\Phi_a^*=\Phi_{\bar{a}}$. Let $\alpha_1,\ldots,\alpha_r$
be the simple roots of $g$ and $\alpha_0=-\theta$ the
negative highest root. The inner products $(\alpha_j,\Phi)$
appearing in the interaction Lagrangian take the form
$(\alpha_j,\Phi)=\sum_a(\gamma_j)_a\Phi_a =:
\gamma_j\cdot \Phi$, where $(\gamma_j)_a =
(\gamma_j)_{\bar{a}}^*$ can be interpreted as the
components of complex simple roots. They are related to the
Frobenius-Perron eigenvector $q_a^{(1)}$ of the Cartan matrix
(c.f. appendix A). The Lagrangian becomes
\be
\cL = \frac{1}{2} \sum_a
\d_{\mu}\Phi_a \d^{\mu}\Phi_{\bar{a}} - \sum_{N\geq 0}
\frac{1}{N!} V_N[\Phi]\;.
\ee
The vertices are
\bas\nspace
&&V_N=\sum_{a_1,\ldots,a_N} \lb_{a_1\ldots a_N}
 :\Phi_{a_1}\ldots\Phi_{a_N}:\;,
\eas
where $:\;\;:$ denotes some normal ordering and
\be
\lb_{a_1\ldots a_N}= m^2\beta^{2(N-2)}
\sum_{a_1,\ldots,a_N} a_j
(\gamma)_{a_1}\ldots (\gamma_j)_{a_N} \;,
\ee
if $a_j$ are the Dynkin labels (\cite{Kac}, p. 54).
In particular, $V_0=h$ (Coxeter number), $V_1=0$ and
from
\bas
\nspace && V_2=2 h m^2 \sum_a q_a^{(1)}q_{\bar{a}}^{(1)}
\Phi_a\Phi_{\bar{a}}\;,
\eas
one reads off the classical masses $m_a=m\sqrt{2h}q_a^{(1)}$.
For $g=A_r$ one finds explicitly,
\ba
&& T_a=\frac{1}{2\sqrt{r+1}\sin\frac{a\pi}{r+1}}
\sum_{j=1}^r(\omega^{aj}-1)\alpha_j\;, \nonum
&& T_a^{\dagger} =T_{\bar{a}},\;\;\bar{a}=r+1-a,\bspace\;\;\;
\Tr(T_aT_b)=\delta_{a,\bar{b}}\;, \nonum
&& (\gamma_j)_a =\frac{2}{\sqrt{r+1}} \sin\frac{a\pi}{r+1}\;
\omega^{aj}\;,\sspace m_a =2m\sin\frac{a\pi}{r+1}\;.
\ea
\newsubsection{Tadpole function and renormalization group}
Next we prepare the set-up for the renormalized perturbation
theory. The free propagator is
\be
\Delta_{ab}(x|m_a)=\delta_{a,\bar{b}}\;i\,\bra 0|T
\Phi_a(x)\Phi_b(0)|0\ket =\delta_{a,\bar{b}}\int
\frac{d^2p}{(2\pi)^2} \frac{e^{ip\cdot x}}%
{p^2-m_a^2 +i\epsilon}\;,
\ee
For spacelike distances $x^2<0$ the evaluation results in
a Bessel function, from which one can read off the
asymptotics
\ba \nspace
\Delta_{ab}(x|m_a) &=& \delta_{a,\bar{b}}\,\frac{1}{2\pi}
                        K_0(m_a|x|) \nonum
                   &=& -\delta_{a,\bar{b}}\,
                   \frac{1}{2\pi}\left(
                   \ln\frac{1}{2}m_a|x| +\gamma\right)
                   +o(|x|)\;,\sspace |x|\ra 0\;,
\ea
where $|x|=\sqrt{-x_{\mu}x^{\mu}}$ and $\gamma$ is the
Euler constant. Since the models have a mass gap, UV
regularization is sufficient and is done by means of a
momentum cutoff $\Lambda$. For example, one can use
$\Delta_{ab}(x|m_a,\Lambda) = \Delta_{ab}(x|m_a) -
\Delta_{ab}(x|\Lambda)$ or
$\Delta_{ab}(x|m_a,\Lambda) = \delta_{a\bar{b}}
\int_{-\Lambda}^{\Lambda}\frac{dk}{4\pi}
\frac{e^{-ikx}}{\sqrt{k^2 +m_a^2}}$ as the regularized
free propoagator.

Renormalization is done by normal ordering w.r.t. a
normal ordering mass $M$. Let $\cF$ be the vector space
of local field operators and define $:\;\;:_M\;:\cF\rra\cF$
to be a linear map satisfying
\be
:\exp\left(\lb\gamma\cdot\Phi\right):_M\,=\,
\exp\left(\lb\gamma\cdot\Phi-\frac{1}{2}\lb^2
\bra (\gamma\cdot\Phi)^2\ket_M \right)\;,
\ee
where $\lb\in\C$ and $\bra (\gamma\cdot\Phi)^2\ket_M =
\gamma_a\gamma_b\Delta_{ab}(0|M,\Lambda)$ is the contraction
function. It is easy to see that (2.6) specifies $:\;\;:_M$
completely. In particular, monomials of the form
$:\Phi_{a_1}\ldots \Phi_{a_N}:_M$ get mapped onto generalized
Hermite polynomials in $\Phi_{a_1},\ldots, \Phi_{a_N}$.
Let $\cO$ be a local operator and let $\bra\;\;\ket_m$ be
the expectation value evaluated by means of Wicks theorem
and the free regularized propagator $\Delta_{ab}
(x|m_a,\Lambda)$. (We make use of the fact that for simply
laced ATs all -- classical and quantum corrected -- masses
are proportional to a single mass scale $m$, which is used
to label $\bra\;\;\ket_m$.) The construction then is designed
s.t. all Greens functions
\bas
\nspace && G_{a_1\ldots a_n}^{\cO}(y_1,\ldots,y_n;x):=
\bra\,T\Phi_{a_1}(y_1)\ldots \Phi_{a_n}(y_n)\,
:\cO(x):_M\, \ket_m
\eas
are finite as $\Lambda\ra \infty$. Even after removal of the
cutoff the theory contains two mass scales $m$ and $M$. One
can also consider $m$ and the dimensionless ratio $M/m$ as
parameters, and put the latter equal to any function
$f(\beta)$ of the coupling constant. We will refer to a
choice of the pair $(m,\,M/m=f(\beta))$ as a choice of the
{\em renormalization scheme}. Interaction Lagrangians which
satisfy
\bas \nspace
&& m_1^2\;\sum_{j=0}^r :e^{\beta\gamma_j\cdot\Phi} :_{M_1} =
m_2^2\;\sum_{j=0}^r :e^{\beta\gamma_j\cdot\Phi} :_{M_2}
\eas
define equivalent renormalized perturbation theories.
{}From (2.6) one checks that this holds iff
\be
m_1(M_1)^{\beta^2/4\pi} = m_2(M_2)^{\beta^2/4\pi} \;.
\ee
The relation (2.7) defines the normal ordering
renormalization group (RG). All physical quantities
should be RG invariant, i.e. should be annihilated by the
differential operator $M\dd{M}-\frac{\beta^2}{4\pi}\dd{m}$.

A major result about the renormalized perturbation
theory is that the dependence on $M/m$ in all Greens functions
enters only through a universal function $T(M/m,\beta)$,
which does not depend on the Greens function considered.
Its exponential is defined through the relation
\bas \nspace
&& e^T=\frac{\beta^2}{m^2 h}\;\bra :V[\Phi](x):_M \ket_m\;,
\eas
where $V[\Phi]$ is the interaction Lagrangian. The
`tadpole function' $T$ itself can be interpreted (and
calculated) as the sum of all connected vacuum diagrams.
To lowest order
\be
T\left(\frac{M}{m},\beta\right) = \frac{\beta^2}{m^2 h}
\sum_{a=1}^r\frac{m_a^2}{4\pi}\ln\left(\frac{M}{m_a}\right)
+ o(\beta^4)\;.
\ee
{}From the previous remarks it follows then that different
renormalization schemes will affect Greens functions
only through different choices for the function
\bas \nspace
&& T_f(\beta) := T\left(\frac{M}{m} =f(\beta),\beta\right)\;.
\eas
For example, one can choose a scheme for which the lowest
order contribution (2.8) (the `fundamental tadpole')
vanishes: Define
\bas\nspace
&& \ln\xi := \frac{1}{2h}\sum_a \left(\frac{m_a}{m}\right)^2
\ln\frac{m_a}{m}\;.
\eas
{}From $\sum_a m^2_a = 2h m^2$ one checks
\bas\nspace
&& T\left(\frac{M}{m},\beta\right)=
\frac{\beta^2}{2\pi}\ln\frac{M}{m\xi} +o(\beta^4)\;,
\eas
so that the choice $M/m =\xi$ kills the fundamental
tadpole. Another preferred scheme is to choose
$M/m=\fh(\beta)$  by the implicit function
theorem s.t. $T(\fh(\beta),\beta)$ vanishes
identically (for a certain range in $\beta$). Of course,
in practice the function  $T(M/m,\beta)$ -- and hence
the defining equation for $\fh$ -- will not be known to
all orders. Theoretically, however, the scheme $(m,\fh)$
is distinguished in that the Greens functions $\Gh$ in this
scheme are obtained by deleting {\em all} tadpole diagrams
from the set of diagrams defining the same Greens function
$G$ in a generic scheme. Moreover, once $\Gh$ is known, the
dependence on the tadpoles -- and hence on $M/m$ -- can be
restored through the formula \cite{DDeV}
\be
G =\exp\left(T\; m^2\dd{m^2}\right)\; \Gh \;.
\ee
The fact that this `tadpole dressing' can be described in
closed form is a consequence of the exponential nature of
the interaction. For the same reason also the $M/m$ --
dependence of $T(M/m,\beta)$ can be found explicitely%
\cite{DDeV}
\be
T\left(\frac{M}{m},\,\beta\right)=
\frac{\beta^2/2\pi}{1+\beta^2/4\pi} \left[
\ln\frac{M}{\xi m} + \frac{2\pi}{\beta^2}
\widehat{T}(\beta) \right]\;,
\ee
where $\widehat{T}(\beta)$ is a function of the coupling
constant alone. An important consequence of (2.10) is
that the combination $m^2 e^T$ is RG invariant i.e.
\be
m_1^2 \;e^{T\left(M_1/m_1,\,\beta\right)} =
m_2^2 \;e^{T\left(M_2/m_2,\,\beta\right)}\;,
\ee
if $(m_1,M_1)$ and $(m_2,M_2)$ are related by (2.7).
In fact, the combination $m^2 e^T$ essentially defines
the physical mass scale and hence better should be RG
invariant.

In preparation of section 4 let us show in detail that
the physical masses and the wave function renormalizations
are RG invariant. Beyond the tree level the fields $\Phi_a$
may develop a non-zero vacuum expectation value $\lb_a$.
(For $g=A_r$ the $\lb_a$'s vanish to all orders due to
$Z_{r+1}$-invariance). Thus define the wave function
renormalization by
\bas \nspace
&& \bra 0|\Phi_a(x) -\lb_a|b(p)\ket_m = \sqrt{Z_a}\,
\delta_{a,\bar{b}}\, e^{-ip\cdot x}\;,\\
&& \bra 0|\Phi_a(x) -\lb_a|0\ket_m = 0 \;.
\eas
The shift $\Phi_a\ra\Phi_a -\lb_a$, when inserted into
the Lagrangian (1.1) affects only the 1-point functions,
and is irrelevant for the stability of the classical mass
ratios under quantum corrections. This stability also
allows one to choose the matrix of wave function
renormalizations diagonal. The exact propagator becomes
\be
G_{ab}(p^2) =\frac{(Z_a Z_b)^{-1/2}}%
{(p^2- m_a^2)\delta_{a\bar{b}} -\Sigma_{ab}(p^2) }\;,
\ee
where $\Sigma_{ab}(p^2)$ is the $\Phi_a\Phi_b$ self energy
i.e. $-i\Sigma_{ab}(p^2)$ is the sum of all 1 PI graphs
with external legs $a,b$. The physical mass is defined
as the position of the pole in (2.12). On dimensional grounds
$G_{ab}(p^2)$ is of the form
\bas \nspace
&& G_{ab}(p^2) =\frac{1}{m^2}\, g_{ab}
\left(\frac{p^2}{m^2},\,\frac{M^2}{m^2},\,\beta\right)\;.
\eas
On the other hand the tadpole-free part $\Gh_{ab}(p^2)$
does not depend on $M$, so that
\bas \nspace
&& \Gh_{ab}(p^2) =\frac{1}{m^2}\, \gh_{ab}
\left(\frac{p^2}{m^2},\,\beta\right)\;.
\eas
{}From the tadpole dressing formula (2.9) one finds the
relation
\be
G_{ab}(p^2) =e^{-T}\,\Gh_{ab}(e^{-T}p^2)\;.
\ee
Let $\mh_a^2$ denote the position of the pole in
$\Gh_{a\bar{a}}(p^2)$. The stability of the mass ratios
under quantum corrections implies
\be
\mh_a = K(\beta)\, m\sqrt{2h}\, q_a^{(1)}\;,
\ee
for some function $K(\beta)$, which on dimensional grounds
is a function of the coupling constant alone. The pole in
$G_{ab}(p^2)$ is at $p^2 =(m_a^2)_{phys}$,
\be
(m_a^2)_{phys}= \mh_a^2\,e^T\;,
\ee
which by (2.11) is indeed RG invariant. To proceed, solve
(2.12) for the self energy i.e.
\bas \nspace
&& \Sigma_{ab}(p^2) =[p^2 -m_a^2]\delta_{a\bar{b}} -
(Z_aZ_b)^{-1/2}[G(p^2)^{-1}]_{ab}\;.
\eas
{}From (2.9) one obtains the tadpole dressing relation
for $\Sigma_{ab}$,
\bas\nspace
&&\Sigma_{ab}(p^2) =m_a^2\,\delta_{a\bar{b}} (e^T-1) +
e^T\,\widehat{\Sigma}_{ab}(p^2e^{-T})\;.
\eas
Together,
\ba
Z_a^{-1} &=&1 - \left.
         \frac{\d\Sigma_{a\bar{a}}(p^2)}{\d p^2}\right|_%
         {p^2 = (m_a^2)_{phys}} \nonum
         &=&1 -e^T\dd{p^2}\left.
         \widehat{\Sigma}_{a\bar{a}}(p^2 e^{-T})\right|_%
         {p^2 = (m_a^2)_{phys}}\;.
\ea
{}From the definition of the tadpole-free parts, and on
dimensional grounds the quantity
$e^T\dd{p^2}\widehat{\Sigma}_{a\bar{a}}(p^2 e^{-T})$
is of the form $S_a\left(\frac{e^{-T}p^2}{m^2},\,
\beta\right)$ for some functions $S_a:\R^2\ra \C$.
 Thus, when evaluated at
$p^2 = (m_a^2)_{phys}$, the $(M,m)$-dependence drops out,
leaving a scheme-independent function of $\beta$ alone.
Guided by similar results in the Sine-Gordon model
\cite{KarWeisz}, the exact expressions for $Z_a(\beta)$
were found in \cite{DDeV}.

\newsubsection{Bootstrap S-matrix and minimal form factor}
As a consequence of the presence of higher order
conservation laws the S-matrix of an integrable theory
factorizes into a product of two particle ones and the
set of in- and outgoing particle momenta is preserved
separately with only the assignment to particles permuted.
The S-matrix element $S_{ab}(\theta)$ for the scattering
of particles $a$ and $b$ is meromorphic in $\theta=
\theta_a-\theta_b$ with period $2\pi i$. The physical
sheet corresponds to $0\leq \mbox{Im}\theta <\pi$.
On very general grounds $S_{ab}(\theta)$ is subject
to a number of functional equations, collectively
referred to as `bootstrap equations'\cite{ZZ}.
For a theory without genuine particle degeneracies
(i.e without multiplets transforming under
a continuous symmetry) these take the simple form
\setcounter{abc}{1}
\renewcommand{\theequation}%
{\thesection.\arabic{equation}\alph{abc}}
\ba
&& S_{ab}(\theta)=S_{ba}(\theta) =S_{ab}(-\theta)^{-1}=
S_{ab}^{*}(-\theta^*)\\
\addtocounter{equation}{-1}
\addtocounter{abc}{1}
&& S_{ab}(i\pi -\theta) = S_{a\bar{b}}(\theta)\\
\addtocounter{equation}{-1}
\addtocounter{abc}{1}
&& S_{da}(\theta +i\eta(a))\,S_{db}(\theta + i\eta(b))\,
S_{dc}(\theta +i\eta(c)) =1 \;.
\ea
\renewcommand{\theequation}{\thesection.\arabic{equation}}
The first equation expresses hermitian analyticity and
(formal) unitarity. The second equation implements crossing
invariance and the third one is the bootstrap equation
proper. The charge conjugation operation $a\rightarrow \bar{a}$
is defined in appendix B. $\eta(l),\; l=a,b,c$ are the imaginary
rapidities of the particles $a,b,c$. They are related to
the conventional fusing angles
$U^{\bar{a}}_{bc}=\pi -\bar{U}^a_{bc}$ etc. by
\ba
&& \eta(a) = -\frac{1}{3}(\bar{U}^b_{ac}- \bar{U}^c_{ba})
\nonumber\\
&& \eta(b) = -\frac{1}{3}(\bar{U}^c_{ba}-
                \bar{U}^a_{cb} -2\pi)\nonumber\\
&& \eta(c) = -\frac{1}{3}(\bar{U}^a_{cb}-
                \bar{U}^b_{ac} +2\pi)\;,
\ea
where $\bar{U}^a_{bc}+\bar{U}^b_{ac}+\bar{U}^c_{ab}=\pi$.
Using (2.17.a,b) and suitably redefining $\theta$ yields
the bootstrap equation in the usual form
$S_{d\bar{c}}(\theta)=S_{da}(\theta-i\bar{U}^b_{ac})
S_{db}(\theta +i\bar{U}^a_{bc})$.

For every simple Lie algebra $g$ there exists a minimal
parameter free solution to these equations\cite{AFZ}.
Using the notation of appendix A it is written
as \cite{Dor}
\be
S_{ab}^{min}(\theta) = \prod_{p=1}^{h}\left( 2p +
\frac{c(a)-c(b)}{2}\right)_{\theta}%
^{(\lambda_a,\Omega^{-p}\gamma_b)}\;,
\ee
where
\be
(\mu)_{\theta}
     := \frac{\sinh(\frac{\theta}{2}+\frac{i\pi\mu}{2h})}
             {\sinh(\frac{\theta}{2}-\frac{i\pi\mu}{2h})}\;.
\ee
If $g$ is simply laced (2.19) in addition also has the
required meromorphy. We henceforth take $g$ to be simply
laced. The S-matrix for real coupling affine Toda theories
is of the form $S_{ab}(\theta)= f_{ab}(\theta)\,
S_{ab}^{min}(\theta)$, where $f_{ab}(\theta)$ carries the
dependence on the
coupling constant s.t. $S_{ab}(\theta)\Big|_{\beta=0}=1$.
Further requirements on $f$ are that it should not
introduce further poles into the physical strip for $\beta>0$
(as the particle content of the theory is already
encoded in $S_{ab}^{min}(\theta)$) and that the signs of
the residues of poles of $S_{ab}^{min}(\theta)$ should
be unchanged after multiplication by $f$. An expression
consistent with these requirements as well as various
checks in perturbation theory is
\ba
&& S_{ab}(\theta) =\prod_{p=1}^{h}
 \left[
 \frac{\left(2p + \frac{c(a)-c(b)}{2}\right)_{\theta}}
      {\left(2p + \frac{c(a)-c(b)}{2}+B \right)_{\theta}}
 \right]^{(\lambda_a,\Omega^{-p}\gamma_b)}\;.
\ea
The parameter $B$ is not fixed by the bootstap equations
but all results known are consistent with
\be
B=\frac{\beta^2/2\pi}{1+\beta^2/4\pi}
\ee
Thus, $B$ plays the role of an `effective coupling' and
contains non-perturbative information.
The fact that (2.19), (2.21) are indeed
solutions to (2.17) will be seen below.

For a local operator ${\cal O}(t,x)$ the form factor
on an asymptotic $N$-particle state is defined by
\be
F^{\cal O}_{a_n\ldots a_1}(\theta_i-\theta_j) =
\bra \Omega|{\cal O}(0)
|a_n(\theta_n),\ldots,a_1(\theta_1)\ket\;,
\ee
where $|\Omega\ket$ is the physical vacuum. From the
Wightman axioms one can argue that these form factors
are subject to a number of functional equations
which conversely are then taken to define the
quantities (2.23) axiomatically. For a given bootstrap
S-matrix these equations have the form of a generalized
Riemann Hilbert problem and the reconstruction of the
correlation functions from them is known as the form
factor bootstrap program\cite{KarWeisz,Smir}. Similar
monodromy problems are known in the context of CFT and
quantum groups, which has been the source of
recent progress\cite{Yang,qKZ,Kyoto2}.
For a diagonal S-matrix the defining equations
for the 2-particle form factor are
\ba
&&F^{\cal O}_{ab}(\theta) = S_{ab}(\theta)\,
F^{\cal O}_{ba}(-\theta) \nonumber\\
&&F^{\cal O}_{ab}(i\pi-\theta) =
e^{2\pi i\omega(\Phi,\cO)}\,
F^{\cal O}_{ba}(i\pi +\theta)\;,
\ea
where $\omega(\Phi,\cO)$ is the relative locality index
of $\Phi$ and $\cO$ (defined via the monodromy of their
operator product).
Moreover $F^{\cal O}_{ab}(\theta)$ is required
to be meromorphic on the physical strip
$0\leq \mbox{Im}\,\theta \leq \pi$ with possible
poles and zeros only on the imaginary axis and
$F^{\cal O}_{ab}(\theta)= o(e^{e^{|\theta|}})$ for
$|\mbox{Re}\,\theta| \rightarrow \infty$. The
solution to (2.24) is then uniquely determined by the
position of the poles and zeros up to a normalization
constant. It can be written as
\be
F^{\cal O}_{ab}(\theta)=
K^{\cal O}_{ab}(\theta)\,
F^{min}_{ab}(\theta)\;,
\ee
where $F^{min}_{ab}(\theta)$ is analytic in
$0\leq \mbox{Im}\,\theta \leq 2\pi$ and is normalized
as $F^{min}_{ab}(i\pi) =1$. The pole factor
$K^{\cal O}_{ab}(\theta)$ carries the dependence on
the local operator and has trivial monodromy
$K(\theta)=K(-\theta)=K(2\pi i +\theta)$.

Suppose that the S-matrix allows for an integral
representation of the form
\be
S_{ab}(\theta) = e^{i\delta_{ab}(\theta)}\;,
\sspace \delta_{ab}(\theta) = \int_{0}^{\infty}
\frac{dt}{t}\, h_{ab}(t)\, \sin\frac{h\theta}{\pi} t\;,
\ee
 where $\delta_{ab}(\theta)$ is the scattering phase and
$h_{ab}(t)=h_{ba}(t)$ is real.
The minimal form factor is then given by\cite{KarWeisz}
\be
F^{min}_{ab}(\theta) = e^{f_{ab}(\theta)}\;,
\sspace f_{ab}(\theta) = \int_{0}^{\infty}
\frac{dt}{t}\, h_{ab}(t)\,
\frac{\sin^2(i\pi-\theta)\frac{ht}{2\pi}}{\sh\,ht}\nonum
\ee
with real and imaginary parts
\ba
&& \mbox{Re}\,f_{ab}(\theta) =\int_{0}^{\infty}
\frac{dt}{t}\frac{h_{ab}(t)}{\sh\,ht}
\left( 1- \mbox{ch}\,ht\,
\cos\frac{h\theta}{\pi}t \right)\nonumber\\
&& \mbox{Im}\,f_{ab}(\theta) =
\frac{1}{2}\delta_{ab}(\theta)\;.
\ea
For the S-matrices (2.19), (2.21) the relation
\be
(\mu)_{\theta}=
-\exp\left\{-2i\int_{0}^{\infty}\frac{dt}{t}
\frac{\sh(\mu -h)t}{\sh\,ht}\,
\sin\frac{h\theta}{\pi}t \right\}\;,
\bspace |\mbox{Im}\,\theta|<\frac{\pi\mu}{h}
\ee
leads to integral representations of the form (2.26),
(2.27). For (2.21) one finds
\be
h_{ab}(t) = \frac{2h\, \sh\,\frac{tB}{2}\,
                 \sh\,\frac{t}{2}(2-B) }{\sh(ht)\,\sh\,t}\;
\left(q_{\bar{a}}(t)\,q_b(t)+q_{\bar{b}}(t)\,q_a(t)\right)\;,
\ee
and a similar expression for (2.19). The functions $q_a(t),\;
a=1,\ldots ,r$ are $2\pi i$-periodic functions in $t$ defined
by $q_a\left(\frac{i\pi n}{h}\right) = iq_a^{(n)},\;
n\in E$, where $q_a^{(n)}$ are the eigenvectors (A.5)
of the Cartan matrix. The integral representations
(2.26), (2.27) are also convenient to derive power series
expansions in $e^{\pm n\theta}$ by complex contour
deformation. For the scattering phase one finds
\ba
&& \delta_{ab}(\theta) = \pm \delta_{ab}^{<>}(\theta)
= \sum_{n\in E}\frac{1}{n}\; D_n\; q^{(n)}_a\,q^{(n)}_b
e^{\pm n\theta} \;,
\bspace \mbox{Re}\,\theta <> 0
\ea
The notations are $\theta_n =\pi n/h$ and
\be
D_n =  \frac{4h\,\sin\frac{\theta_n}{2}B
                           \sin\frac{\theta_n}{2}(2-B)
                      }{\sin \theta_n}
    = \beta^2 \, n\left( 1 - \frac{\beta^2}{4\pi}n\,
      \cot \theta_n\right) + o(\beta^6)\;,
\ee
where for later use we displayed the semi classical limit.
As a check one can also obtain (2.32) directly from (2.21)
using the formulae of appendix A.  Similar expansions
exist for the function elements of $\mbox{Re}\,f_{ab}(\theta)$.
Due to the second order pole in $\sh\,ht$ the coefficients
will also carry a linear $\theta$ dependence. We shall
not need there explicit form.

\noindent{\em Remark:}
\noindent $i.$ The corresponding expression (2.31) for the
 minimal S-matrix is obtained by replacing $D_n$ by
$-2h\cot \theta_n$. In addition, there is also a
zero mode contribution to $\delta_{ab}^{<>}(\theta)$ with
value  $\pm \pi(1_{ab}-2(a^{-1})_{ab})$,
where $1$ denotes the unit matrix and $a^{-1}$ is the
inverse of the Cartan matrix.  This implies
$-\frac{1}{2\pi i}\Big[\ln S_{ab}^{min}(\theta)
\Big]_{\theta =-\infty}^{\theta=\infty} = 2a^{-1}-1$,
which has been obtained previously(e.g. \cite{Dor}).
The absence of a zero mode in $\delta_{ab}^{<>}(\theta)$,
of course, is required for $S_{ab}(\theta)|_{\beta=0}=1$.

\noindent $ii.$ Given the additional information that the
series expansion (2.31) indeed defines a meromorphic
function with the correct pole structure\cite{Olive}, one
can also easily verify that it provides a solution to the
bootstrap equations (2.17). From $q_{\bar{a}}^{(n)}=
(-)^{n+1}q_{a}^{(n)}$ one has by analytic continuation
\be
\delta_{\bar{a}b}(i\pi -\theta) = \delta_{ab}(\theta)=
-\delta_{ab}(-\theta)\;,
\ee
which implements equations (2.17.a,b) for the S-matrix.
In terms of the scattering phase the bootstrap
eqn. is equivalent to
\be
\sum_{l=a,b,c}\delta_{dl}(\theta+i\eta(l)) =0\;,
\ee
which follows from (A.13).
\pagebreak
\newsection{Definition of the quantum spectrum}
We will consider ATs in $\d_+$-lightcone dynamics.
Such a lightcone quantization simplifies the structure
of the conserved currents, but also introduces some
complications. See \cite{MN3,MN2} for a detailed discussion.
Being complicated composite operators, the construction
of the quantum conserved currents is a difficult problem
and except for some sample calculations \cite{Niss}
has not been successful in earlier attempts. The major
result in lightcone dynamics is that the construction
of the quantum conserved currents can be mapped onto a
conformal field theory (CFT) problem. In the CFT context,
the existence of infinitely many quantum conserved
currents -- and hence the integrability of the
quantum ATs -- can be proved\cite{MN1,FF}. The construction
can also be rephrased in algorithmic terms and,
at least in principle allows one to explicitly construct
the quantum conserved currents. To specify the structure
of the quantum conserved currents it therefore suffices to
describe the mapping onto (and back from) the CFT problem.

\newsubsection{Construction of the quantum conserved
currents}
Recall that in lightcone dynamics the interaction
lagrangian plays the role of the hamiltonian. For ATs
thus
\be
H[\Phi] = \int  dx^- V[\Phi] =
\frac{m^2}{\beta^2}\sum_{j=0}^r\int dx^-
\left[ :e^{\beta\gamma_j\cdot \Phi}:_M -1\right]
\ee
is the normal ordered lightcone hamiltonian. By definition
it is RG invariant. We wish to construct quantum versions
of the local conserved charges $\In_{\pm},\;n\in E$
present in the classical theory. A quantum conserved
charge is defined to be a functional $I[\Phi]$ commuting
with the Hamiltonian (3.1). A local conserved charge
$I[\Phi] =\int dx^- J[\Phi]$ is supposed to arise by
integration from some density $J[\Phi]$, which is a
normal ordered differential polynomial in $\dmiphi$
and $e^{\beta\gamma_j\cdot \Phi}$. In terms of the
densities $V[\Phi]$ and $J[\Phi]$ the condition reads
\be
[V(x)\,,\,J(y)] =\delta(x-y) \d Q^{(1)}(y) +
\sum_{k\geq 2}\delta^{(k)}(x-y)\,Q^{(k)}(y)\;,
\ee
for some local operators $Q^{(k)},\;k\geq 1$. Here the
commutator is taken at equal `$x^+$--time' and $x= x^-,\;
y=y^-$, etc.. In the classical theory the equation (3.2)
(with the commutator replaced by Poisson brackets)
is known to possess two infinite sequences of solutions
$\Jn_{\pm}[\phi]$, one pair for each affine exponent
$n\in E$ (including multiplicities). For the purposes here
it suffices to consider the densities $\Jn[\dmiphi]:=
\Jn_-[\dmiphi]$, which are differential polynomials in
$\dmiphi$. For such differential polynomials the
commutator (3.2) in the quantum theory is equivalent
to an operator product expansion of the form
\be
V(x)\,J(y) =\frac{1}{x-y+i\epsilon}\d Q^{(1)}(y) +
\sum_{k\geq 2}
\frac{\widetilde{Q}^{(k)}(y)}{(x-y+i\epsilon)^k}\;.
\ee
{}From the distributional formula
$(x+i\epsilon)^{-k} -(x-i\epsilon)^{-k} =
\frac{2\pi i (-)^k}{(k-1)!}\,\delta^{(k-1)}(x)$ one
recovers the commutator (3.2) (with $Q^{(k)}$ related to
$\widetilde{Q}^{(l)},\; l\geq k$). Usually, the calculation
of an operator product expansion in a massive QFT is
not an attractive task. In the case at hand, however,
we can take advantage of the fact that the operator
$V[\Phi]$ is RG invariant. In particular, we can choose
the scheme $(m,\fh)$ defined in section 2.1
to find solutions to (3.2). In this scheme
\bas \nspace
&& 1= e^T = \frac{\beta^2}{h m^2}\bra :V[\Phi]:_M\ket_m\;.
\eas
Further, the quantum masses coincide with the classical
masses $\mh_a =m_a$, and the exact propagator coincides
with the free propagator. Thus, one can simply use the
short distance behaviour of the free propagator (2.5) to
calculate the singular part of the OPE. Moreover, since
(for $J=J[\dmiphi]\,$) all contractions in (3.2) involve
at least one derivative field, even the dependence on
the classical masses drops out. The singular part of the
resulting OPE will be form-identical to the one obtained
by using the contraction function
\bas \nspace
&& \Phi_a(x)\,\Phi_b(y) =-\delta_{a\bar{b}}\frac{1}{2\pi}
\ln(x+i\epsilon)\;.
\eas
But this is (up to a normalization) just the contraction
function that one would use for free massless fields in a
Minkowski space CFT. Hence if one is able to find solutions
to (3.2) in the CFT, the normal ordered operators $J[\dmiphi]$
will also have an interpretation in the massive QFT: They
define quantum conserved densities $J[\dmiphi]$ of the
ATs in the particular renormalization scheme $(m,\fh(\beta))$.
It remains to solve the problem in the CFT context.
The result \cite{MN1,FF} is that to each classical conserved
density $(\Jn[\dmiphi])_{class},\;n\in E$, there exists
a unique normal ordered quantum operator $\Jn[\dmiphi]$
solving (3.2). In the CFT this can be shown for all affine
Lie algebras. Since in non-simply laced AT's
the stability of the mass ratios under quantum corrections
ceases to hold, the preferred scheme $(m,\fh(\beta))$
does not exist in these cases. Therefore, for non-simply
laced Lie algebras, the CFT solutions to (3.2) presumably
have no direct significance in AT's. In principle
(returning to the simply laced cases) one can also use
the defining relation (3.2) to explicitely construct the
conserved densities $\Jn[\dmiphi]$ by fixing the
coefficients in an in an appropriate Ansatz. The
coefficients turn out to be constrained by overdetermined
linear systems (of rapidly increasing size in $n$).
The existence theorem quoted guarantees that these
overdetermined linear systems can always be solved, but
doing it explicitely soon becomes inconvenient. A more
economical way to construct the conserved charges
explicitely is to exploit the relation to $W$-algebras
because then part of the constraints are
built in\cite{KM, MN1}.
For later use we quote some examples for
the $A$-series. We normalize the $\Jn$'s according to
the form in which they appear in the (quantum) Miura
transformation. That is
\be
\Jn[\d\Phi]=(-\beta)^{n+1} s_{n+1}[(h_j,\d\Phi)]+\ldots\;,
\bspace 1\leq n\leq r\;,
\ee
where $h_0,\ldots, h_r$ are the weights of the $r+1$-%
dimensional fundamental representation and $s_n$ is the
totally symmetric polynomial of order $n$ in $r+1$ variables.
The normalization of the higher quantum conserved densities
is fixed with reference to their classical counterparts
(c.f. section 4.1). Using the transformation
\be
(h_j,\Phi)=\frac{i}{\sqrt{r+1}}\sum_{a=1}^r
\omega^{-a/2}\omega^{aj}\;\Phi_a\;,
\ee
one finds
\ba
\beta^{-2}J^{(1)} & = &
\frac{1}{2}\sum_a \d\Phi_a\d\Phi_{\bar{a}}\;,
\nonum
\beta^{-3}J^{(2)} & = &-i\frac{r\sqrt{r+1}}{18}
\sum_{a+b+c\,\equiv\, 0}
\omega^{(a+b+c)/2}\;\d\Phi_a\d\Phi_b\d\Phi_c  \nonum
&& +\frac{i}{2}\left(\frac{1}{\beta}
+\frac{\beta}{4\pi}\right)
 \sum_a \cot\frac{a\pi}{r+1}\; \d^2\Phi_a\d\Phi_{\bar{a}}
\ea
Similarly $\Jn,\;1\leq n\leq r$ can be obtained
from the Miura transformation and the higher conserved
densities are then calculated by means of the
procedure outlined. In general they will not be known
explicitely. Remarkably, at least the quadratic part
$J^{(n,2)}$ of the $\Jn$'s can be given in closed form
\ba
\beta^{-(n+1)} J^{(n,2)}& = &\beta^{-n+1}
 \left(1+ \frac{\beta^2}{4\pi} +o(\beta^4)\right)\;
\sum_a c_a^{(n)}\,\d^n\Phi_a\,\d\Phi_{\bar{a}} \;,\nonum
&& c_a^{(n)} = i^{n-1}\frac{\sin\frac{a n\pi}{r+1}}%
               {\left(2\sin\frac{a \pi}{r+1}\right)^n}\;.
\ea
We defer the derivation of (3.7) to section 4.1.
\newsubsection{The current algebra}
The results of the last subsection allow one to deduce
some results on the structure of generalized current
algebras in AT's. In 2-dim.~Minkowski space consider
a current algebra in $\dpl$-lightcone dynamics
of the following form.
\ba
&&\nspace  [P^{(i)}(f)\;,\;P^{(j)}(g)] =
\sum_{\{k:\Delta_{ijk} \geq 1\}} C^{ij}_k\;
P^{(k)}\Big(p^{\Delta_i\Delta_j}_{\Delta_k}(f,g)\Big) +
D^{ij}\, \omega^{\Delta_i}(f,g)\;.
\ea
The commutator is taken at equal $x^+$-time and
the range of the summation is defined in terms of
$\Delta_{ijk}=\Delta_i +\Delta_j -\Delta_k$, where
$\Delta_i$ is the Lorentz spin of the field
$P^{(i)}$. Here $P^{(i)}(x) := P^{(i)}(0,x),\;x=x^-$
denotes some normal ordered differential polynomial
of the fundamental fields. The structure constants
$C^{ij}_k,\; D^{ij}$ are dynamical parameters of the
theory. The operators are smeared via
\bas
P^{(i)}(f):= \int dx P^{(i)}(x)f(x) \; .
\eas
$f,\;g$ and $p^{\Delta_i\Delta_j}_{\Delta_k}(f,g)$ are
smooth test functions, the latter being given by
\ba
&&\nspace  p^{\Delta_i\Delta_j}_{\Delta_k}(f,g)=
\frac{1}{(d-1)!}\frac{(2\Delta_k -1)!}{(2 \Delta_k +d-2)!}
\sum_{r=0}^{d-1} (-)^r
\left(\!\! \begin{array}{cc} d-1\\r \end{array}\!\! \right)
c_r\, f^{(d-1-r)} g^{(r)} \;,
\ea
where $d= \Delta_{ijk},\; f^{(n)}=(-i\dmi)^{n}f$ and
\bas\nspace
&&  c_r = \frac{(2 \Delta_j -2 -r)!}{(2\Delta_j -d-1)!}
          \frac{(2 \Delta_i -d- 1+r)!}{(2\Delta_i -d-1)!}\;.
\eas
The cocycle is
\bas\nspace
&&\omega^{\Delta_k}(f,g) = \frac{1}{(2\Delta_k -1)!}
\int dx^-f^{(2\Delta_k-1)}g\;.
\eas
In particular, let $P_-=2I^{(1)}_-$ and $K$ denote the
generators of infinitesimal translations and Lorentz
boosts, respectively. Then
\ba
&& [P_-\,,\; P^{(k)}(x)] = -i\d P^{(k)}\nonumber\\
&& [K\,,\; P^{(k)}(x)] = (-ix\d + \Delta_k) P^{(k)}\;.
\ea
We claim that
\begin{itemize}
\item[--] (3.8) is the most general current algebra
possible in a 2-dim.~QFT in lightcone dynamics, which
is compatible with Poincare' invariance. Only the
`structure constants' $C^{ij}_k,\;D^{ij}$ are dynamical
parameters of the theory.
\item[--] The algebra in (3.8) is isomorphic to a
vertex operator algebra or meromorphic CFT (in the
sense of \cite{mCFT}) with the same structure constants.
The isomorphism can be constructed explicitely.
\end{itemize}
For the proof we refer to \cite{Carg,MN3}. To illustrate
the point of the construction consider a scalar
QFT with non-derivative interaction. The $T_{--}$
component of the energy momentum tensor is then
given by $\dmi\Phi\cdot\dmi\Phi$ and the operators
$L_s = \frac{1}{2}\int dx^- (x^-)^{s+1} \dmiphi
\cdot\dmiphi,\;s=0,\;\pm 1$ generate an $so(1,2)$
algebra. By adding suitable total derivative
terms one can achieve that $P^{(k)}[\dmiphi]$, and
hence a suitable basis of normal ordered products
thereof, transforms covariantly under this $so(1,2)$
algebra. This is due to the
fact that in lightcone coordinates Lorentz boosts
$x^+\rightarrow e^{\theta}x^+,\; x^-\rightarrow
e^{-\theta}x^-$ can be viewed as scale transformations
$x^{\pm}\rightarrow \rho x^{\pm}$ with separate scale
factors for each lightcone sector. For either one
of the lightcone sectors (but not for both
simultaneously) the action of the Poincare algebra
on the equivalence classes of the field algebra
modulo total derivatives therefore extends to an action
of the (chiral part of the)
finite conformal group $SO(1,2)$. One can then use CFT
techniques to determine the $SO(1,2)$-covariant form of
the commutator $[P^{(k)}(f)\,,\;P^{(l)}(g)]$. In particular,
the mCFT definition of the structure constants
$C^{ij}_k,\;D^{ij}$ implies that, if one of the fields
is the Virasoro generator, only the structure constant
$C^{L,P^k}_{P^k}$ is non-vanishing and equals 2 in
standard normalizations. This guarantees the
consistency of the general formula (3.8) with (3.10)
and gives the correspondence $P_-\ra L_{-1},\;K\ra L_0$
to mCFT generators. To check this, note that
\bas
p^{2j}_j(f,g)= \frac{1}{2}
[(\Delta_j-1)f^{(1)}\,g- g^{(1)}\,f]\;.
\eas
Choosing $f=1, x$, respectively one recovers
the equations (3.10).

Let us also briefly describe the mapping onto a mCFT in
general. For a current of Lorentz spin $\Delta$ in (3.8),
the image in the mCFT is given by
\be
P(x) \longrightarrow P(z)=
\sum_{n\in \Z} P_n\, z^{-n-\Delta}\;,
\ee
where on the r.h.s. $P(z)$ is a quasiprimary field of weight
$\Delta$. The modes are defined by
\be
P(x) =: \left(\frac{2}{1+x^2}\right)^{\Delta} \sum_{n} P_n
\left(\frac{1+ix}{1-ix}\right)^{-n}\;.
\ee
In particular the integral $\int dx P(x)$ is mapped onto
a mCFT quantity lying in the vacuum preserving subalgebra
$P_n,\;|n|\leq \Delta -1$ of the mCFT.
\be
\int dx P(x) \longrightarrow
4\pi\, 2^{-\Delta_i} \sum_{|n|\leq \Delta_i -1}
\left( \begin{array}{c} 2\Delta_i -2 \\
                        | n+ \Delta_i -1|
       \end{array}
\right)
P_n \;.
\ee
By means of a further
transformation one can then achieve that $\int P$ is
mapped onto $P_{\pm(\Delta -1)}$ or even onto a zero mode
$\widetilde{P}_0$ for some modified field $\widetilde{P}(z)$.
See \cite{Carg} for more details.

These results hold for any normal ordered differential
polynomial in the fundamental fields. In particular, the
conserved densities $\Jn =\Jn[\dmiphi]$ are of that type.
We conclude that the conserved densities $\Jn$ together
with all their normal ordered products form a generalized
current algebra of the form (3.8). The $\Jn$'s alone
will in general not  form a closed algebra by themselves.
The existence of the conserved charges is equivalent
to the vanishing of a certain subset of the
structure constants. From (3.8), (3.9) one checks
\be
[\In\,,\,I^{(m)}]=0 \sspace \mbox{iff}\;C_k^{mn}=0\;\;
\forall k :\;\Delta_{mnk}=1\;,
\ee
if $n,m$ refer to the densities $\Jn,\;\Jm$ and
$k$ refers to any composite field $P^{(k)}$ appearing
on the r.h.s. of (3.8). Since the structure constants
in (3.8) coincide with that in the mCFT, this confirms
again that the existence of the conserved charges is
decided on the CFT level.
%
\newsubsection{Definition of the quantum spectrum}
Consider now the eigenvalues of the conserved charges
$I^{(n)}_{\pm}$ on multiparticle states. Let
$|a_N(\theta_N),\ldots ,a_1(\theta_1)\ket$
denote an asymptotic multiparticle state, where
$\theta_i$ is the rapidity of a particle of type
$a_i$. The action of the charges $I^{(n)}_{\pm}$ on
these states is expected to be of the form
\ba
&& I^{(n)}_-|a_N(\theta_N),\ldots ,a_1(\theta_1)\ket
= \sum_{k=1}^{N} I^{(n)}_-(a_k)\; e^{-(\Delta_n-1)\theta_k}
|a_N(\theta_N),\ldots ,a_1(\theta_1)\ket  \nonum
&& I^{(n)}_+|a_N(\theta_N),\ldots ,a_1(\theta_1)\ket
= \sum_{k=1}^{N} I^{(n)}_+(a_k)\; e^{-(\Delta_n-3)\theta_k}
|a_N(\theta_N),\ldots ,a_1(\theta_1)\ket
\ea
Such common eigenstates exist because of $[\Inpl\,,\,\Inmi]=0$.
The factorization of the eigenvalues into a sum of single
particle eigenvalues is a consequence of the locality of
the charges and the approximate independence of the
localized asymptotic wavepackets. We anticipate also
that the single particle eigenvalues $I^{(n)}_{\pm}(a)$
coincide up to a sign
\be
I^{(n)}(a):= I^{(n)}_-(a) = -I^{(n)}_+(a) \;.
\ee
This is a consequence of the current conservation equation
in momentum space \newline $p_+\,\Jnmi(p_+,p_-) +
p_-\,\Jnpl(p_+,p_-)=0$, which holds on the operator level
and a-fortiori relates the matrix elements of
$J^{(n)}_{\pm}(p_+,p_-)$. On the other hand the eigenvalues
$I^{(n)}_{\pm}(a)$ can be calculated from on-shell single
particle matrix elements (c.f. below)
\bas \nspace
&& I^{(n)}_{\pm}(a) \sim \left. \bra a(p)|
J^{(n)}_{\pm}(p_+,0)|a(p)\ket
\right|_{p^2 =(m_a^2)_{phys}}\;.
\eas
This implies $p_+ \Inmi(a) +p_-\Inpl(a) =0$ at
$p^2 =(m_a^2)_{phys}$ and hence (3.16) for rapidity
$\theta =0$.

Therefore it suffices to consider the $\Inmi$ charge
and we simplify the notation by putting $J^{(n)}=
\Jnmi$, $I^{(n)}=\Inmi$ etc.. Consider then in detail the
matrix elements of $I^{(n)}$ between single particle
states of type $a$. We use the standard normalization
transcribed to rapidity variables
\bas \nspace
&& \bra a(p_1)|a(p_2')\ket = 2\pi p_0 \delta(p_1-p_1')\;;
\sspace  \bra a(\theta)|a(\theta')\ket =
4\pi \delta(\theta-\theta') \;.
\eas
By definition
\ba \nspace
&& 4\pi\delta(\theta -\theta')\,e^{-n\theta}\,\In(a)=
\bra a(\theta)| \In | a(\theta')\ket =
\bra a(\theta)| \int dx^- \Jn[\dmiphi]
|a(\theta')\ket \;.
\ea
On the other hand for any local operator $\cO(x)$ an
LSZ reduction yields the relation
\ba
\bra a(p)|\cO(x)| a(q)\ket &=& \lim_%
{p^2 =(m_a^2)_{phys},q^2 =(m_b^2)_{phys}}
\left(\frac{-i}{\sqrt{Z_a}}\right)
\left(\frac{-i}{\sqrt{ Z_{\bar{a}} }}\right) \times \nonum
&&\sspace\times [p^2 -(m_a^2)_{phys}][q^2 -(m_b^2)_{phys}]\;
G_{ab}^{\cO}(-p,q)\;.
\ea
In the case at hand $J^{(n)}(0,x^-)$ is integrated along
$x^-$ and the relevant momentum space Greens function is
defined by
\be
2\pi \delta(p_-+q_-)\,G^{\int\Jn}_{a\bar{a}}(p,q):=
\int d^2y_1d^2y_2\,e^{ip\cdot y_1}\,e^{iq\cdot y_2}\;
G^{\int\Jn}_{a\bar{a}}(y_1,y_2)\;,
\ee
where $p=(p_+,p_-),\;q=(p_+,q_-)$ and
\bas \nspace
&& iG^{\int\Jn}_{a\bar{a}}(y_1,y_2)=
\bra \Omega|\int dx^- \Jn(0,x^+)\,\Phi_a(y_1)
\Phi_{\bar{a}}(y_2)|\Omega\ket\;.
\eas
Since only connected diagrams contribute, the r.h.s.~%
before integration will depend on the differences
$x-y_1,\;x-y_2$ only.
The $dx^-$ integration thus will result in a factor
$2\pi \delta(p_-+q_-)$, which has already been extracted
in the definition of $G^{\int\Jn}_{a\bar{a}}(p,q)$. The
same factor will appear on the r.h.s. of (3.17) when
applying the LSZ formula (3.18) to $\int dx^- J^{(n)}$.
Together
\be
\In(a) = -(Z_aZ_{\bar{a}})^{-1/2}\;
\lim_{p^2 \ra (m_a^2)_{phys}}\,\frac{e^{n\theta}}{2p_-}\;
[p^2 -(m_a^2)_{phys}]^2\, \Gn_a(p)\;,
\ee
where $\Gn_a(p)=
G^{\int\Jn}_{a\bar{a}}(p_+,-p_-;p_+,p_-)$.
The equation (3.20) gives a perturbative definition of the
eigenvalue $\In(a)$. As for the SG model in the super-%
renormalizable regime, one expects the perturbation theory
to have non-zero radius of convergence, so that (3.20) amounts
to an exact, unambigous definition. As a check note that
the Feyman diagrams contributing to $\Gn_a(p)$ will produce
a factor $(p_-)^{n+1}\times [p^2 -(m_a^2)_{phys}]^{-2}\times
(\mbox{dim.less quantity})$. This means for the eigenvalue
$\In(a)\sim m^n\times (\mbox{dim.less quantity})$. In
detail, we claim that $\In(a)$ has the following form
\be
\In(a)=K_n(\beta)\left(\frac{m^2\,e^T}{2\beta^2}
\right)^{n/2}\; q_a^{(n)}\;.
\ee
Here $K_n(\beta)$ is an unknown function of the coupling
constant to be determined (from which we extracted a factor
$(2\beta^2)^{-n/2}$ for later convenience). Further,
$m^2\,e^T$ is the RG invariant combination (2.11) and
$q_a^{(n)}=q_a^{(n+h)}$ are the components of the $r$-th
($n=r\,\mbox{mod}\,h)$ eigenvector of the Cartan matrix.

To prove (3.21), first recall that $\Gn_a(p)$ is the product
of $(p_-)^{n+1}\times [p^2 -(m_a^2)_{phys}]^{-2}$ and a
dimensionless quantity composed of Lorentz scalars. On
dimensional grounds one thus knows
\bas \nspace
&& \frac{e^{n\theta}}{2p_-}\,[p^2 -(m_a^2)_{phys}]^2\;
\Gn_a(p) =m^n\,\gn_a
\left(\frac{p^2}{m^2},\frac{p^2}{M^2},\beta\right)\;,
\eas
for some function $\gn_a :\R^3\ra \C$. Similarly one
defines its tadpole-free counterpart
$\gh^{(n)}_a(p^2/m^2,\beta)$.
As usual, both are related by tadpole dressing
\ba
m^n\,\gn_a(p^2/m^2,\beta) &=&
\exp\left(T\,m^2\dd{m^2}\right)
\left[m^n\,\gh^{(n)}_a(p^2/m^2,\beta)\right] \nonum
                & = & \left(e^T\,m^2\right)^{n/2}\,
                \gh^{(n)}_a(e^{-T}p^2/m^2,\beta)\;.
\ea
This has to be evaluated at $p^2=(m^2_a)_{phys}$. Since
$(m^2_a)_{phys}\sim m^2\,e^T$, the $(M,m)$ dependence
in $\gh^{(n)}$ drops out, leaving  a scheme independent
function $-k_a^{(n)}$ of the coupling alone. Together
\be
\In(a) =\left[(m^2_a)_{phys}\right]^{n/2}
(Z_aZ_{\bar{a}})^{-1/2}\,k_a^{(n)}(\beta)\;.
\ee
The wave function renormalizations $Z_a(\beta)$ have
already been seen to be scheme independent. The
form (3.23) of the eigenvalues therefore implies in
particular that they are scheme independent physical
quantities. The form of $\In(a)$ is further constrained
by the `conserved charge bootstrap'\cite{Z}. This is
to say that the eigenvalue equations (3.15) have to be
compatible with the formation of bound states. In the
notation of appendix A, this enforces the condition
\be
\sum_{l=a,b,c} e^{\pm\eta(l)}\,\In(l) =0\;,
\ee
where $\eta(l)$ is defined in (2.18). This requires
$\In(a) \sim q_a^{(n)}$ i.e. in (3.23)
$(q_a^{(1)})^n\, k_a^{(n)} \sim q_a^{(n)}$. We
conclude that $\In(a)$ has the form (3.21), which is
what we wanted to show.

We note some consequences of the relation (3.21). First,
$\In(a)$ is composed of scheme independent quantities,
and hence is itself scheme independent. Still, both factors
have a slightly different status. $K_n(\beta)$ is a
universal, scheme independent function of $\beta$, to be
determined. The second factor $(m^2\,e^T)^{n/2}$, although
numerically constant under a RG transformation, will be
represented by different functions of $\beta$ in different
schemes. However, once $m^2\,e^T$ is known in one scheme
as a function of $\beta$, it is known in any other: Suppose
two renormalization schemes $M_1/m_1 =f_1(\beta)$ and
$M_2/m_2 =f_2(\beta)$ to be given. Then
\bas \nspace
&& \frac{m_1}{m_2} =
\left(\frac{f_2(\beta)}{f_1(\beta)}\right)^{B/2}\;,
\bspace B=\frac{\beta^2/2\pi}{1+\beta^2/4\pi}\;.
\eas
Suppose further that $e^{T(f_1(\beta),\beta)}
=: T_1(\beta)$ is known explicitely as a function of
$\beta$. By RG invariance $m_1^2\,T_1(\beta)=
m_2^2\,T_2(\beta)$, where $T_2(\beta):=
e^{T(f_2(\beta),\beta)}$. Thus also
\be
T_2(\beta) = \left(\frac{m_1}{m_2}\right)^2\,T_1(\beta)
=\left(\frac{f_2(\beta)}{f_1(\beta)}\right)^B \,T_1(\beta)
\ee
is known explicitely. For the eigenvalues this implies
that if $\In(a)$ is known explicitely as a function of
some bare mass scale $m_1$ and the coupling constant, it
is also known explicitely as a function of any other bare
mass scale $m_2$ and the coupling constant. It remains
to determine the unknown functions $K_n(\beta)$.

\newsection{Perturbative evaluation of $K_n(\beta)$}
In this section we restrict attention to the
$A_r^{(1)}$ series.
\vspace{-1cm}

\newsubsection{The classical limit}
The classical limit corresponds to the tree level diagrams
in (3.20). At tree level $Z_a =1=T$ and since only connected
diagrams enter, only the quadratic part of $\Jn$ will
contribute. Modulo total derivatives $\Jn$ is invariant
under the action of the dihedral group generated by
$\Phi_a \ra \omega^a\Phi_a,\;\Phi_a \ra \Phi_{\bar{a}}$
\cite{MN1}. For the quadratic part
$J^{(n,2)}$ this implies
\be
J^{(n,2)} =\beta^{-(n+1)}\sum_a c_a^{(n)}\,
\d^n\Phi_a\d\Phi_{\bar{a}} \;,
\ee
for some coefficients $c_a^{(n)}$ satisfying
$c_{\bar{a}}^{(n)}=(-)^{n+1}c_a^{(n)}$. Inserting
into (3.19) gives
\bas \nspace
&& G_a^{(n)}(p) =2(-i)^{n+1}\,\beta^{-n+1}
\,c_a^{(n)}\,p_-^{n+1}\,\frac{1}{(p^2 -m_a^2)^2}\;,
\eas
where $m_a =2m\sin\frac{a\pi}{r+1}$ are the classical
masses. From $p_-=m_a e^{-\theta}/\sqrt{2}$ one obtains
\be
\In(a) =(-i)^{n-1} \left(\frac{m_a}{\sqrt{2}}\right)^n\,
\beta^{-n+1}\,c_a^{(n)}\;.
\ee
It remains to determine the coefficients $c_a^{(n)}$.
{}From the defining relations (4.1), (3.2) this is likely
to be formidable. From (4.2) it is also clear that the
calculation of the $c_a^{(n)}$'s essentially amounts
to a direct calculation of the eigenvalues $\In(a)$.
Fortunately, in the classical model the latter is
possible by means of a trace identity, which gives an
expression for the conserved charges in terms of the
scattering data of a given solution
\be
I^{(n)}[solution] = I^{(n)}(scattering\;data)\;,
\sspace n\in E\subset \N\;.
\ee
The solutions relevant here should have the following
features
\begin{itemize}
\item[--] The solutions should be singular solitary wave
solutions i.e. solutions with discrete scattering data,
which describe the propagation of point-like singularities
in the initial data.
\item[--] The classical masses should coincide with the
ones obtained from the quadratic part of the Lagrangian
(1.1).
\item[--] Their trace identity (4.3) should coincide with
the leading order term in $\beta$ of the trace identity
for the breather solutions in the imaginary coupling
model.
\end{itemize}
These requirements parallel the situation in the
Sinh-Gordon model\cite{PP}. Since the trace identity
for the breathers in the imaginary coupling AT's
is known\cite{MN2}, the last condition is the most
convenient one to obtain the result. By construction,
the result will then also meet the second requirement,
and a closer inspection of the corresponding solutions
(defined in terms of their $\tau$-functions) shows that
they indeed describe the propagation of pointlike
singularities in the initial data. We omit the details.
Thus, the trace identity for the breather solutions
in the imaginary coupling model is needed. Such
breather solutions
are parametrized by their type $a\in\{1,\ldots,r\}$
and a triplet of action-angle variables. The triplet of
action variables is $(\theta,\rho,j)$, where $\theta$ is
the rapidity, and $\rho$ and $j\in\{1,\ldots, a\}$ are a
continous and discrete excitation number, respectively.
Conjugate to them is a triplet of angle variables which
does not enter the trace identity. Explicitely, the
trace identity at zero rapidity for a breather of type
$a$ and action variables $(\theta=0,\rho,j)$ is%
\footnote{We use slightly different conventions as
in \cite{MN3,MN2}. The classical mass of a solution
is $m[solution] =2\sqrt{2}I^{(1)}[solution]$.}
\ba \nspace
&& \In[breath_a]= -\frac{r+1}{n}\,(i\beta)^{-(n+1)}
\left(\frac{m}{\sqrt{2}}\right)^n\,\times \nonum
&& \bspace\sspace\times\sin\left(
\frac{\beta^2 n\rho}{r+1}+\frac{\pi n}{r+1}(j-1)
\right) \;\sin\frac{a n \pi}{r+1}\;.
\ea
For $n=1$ the leading term in $\beta$ (after changing
$i\beta$ to $\beta$) should reproduce the classical
masses in (2.3). This fixes the parameters $\rho=1,j=1$
i.e. the singular solitary wave solution of type $a$ in the
real coupling model is in correspondence with the breather
solution of type $a$ and excitation numbers $(\rho,j)=(1,1)$
in the imaginary coupling model. With these parameters fixed
the procedure is repeated for generic $n$. Changing $i\beta$
to $\beta$ in (4.4) and extracting the leading order term
in $\beta$ yields the trace identity at zero rapidity
\be
\In[ssw_a] =\beta^{-n+1}
\left(\frac{m}{\sqrt{2}}\right)^n\,
\sin\frac{an\pi}{r+1}\;,
\ee
for the singular solitary wave solution $ssw_a$ of type
$a$ in the real coupling model. Equation (4.5) gives
the classical eigenvalues searched for
\be
\In(a) =\In[ssw_a]\;.
\ee
In particular, comparing with (4.2) one can read off the
coefficients
\be
c_a^{(n)} =i^{n-1}\frac{ \sin\frac{a n\pi}{r+1} }%
{ \left( 2\sin\frac{a\pi}{r+1}\right)^n }\;.
\ee
defined by (4.1). For $n=1,2$ this is also confirmed by
direct calculation (c.f. (3.6)). On the other hand
comparing (4.6) with the generic form (3.21) one can
read off the classical limit of the unknown function
$K_n(\beta)$.
\be
\left(K_n(\beta)\right)_{class} =
\sqrt{\frac{h}{2}}\beta\;.
\ee
\newsubsection{First order perturbation theory}
To do explicit calculations in PT it is convenient to
rescale the fields $\Phi_a \ra \beta^{-1}\Phi_a$. The
coupling constant $\beta^2$ will then serve as a loop
counting parameter. All vertices $\lb_{a_1\ldots a_N}$
are proportional to $\beta^{-2}$ and the propagator
carries an extra factor of $\beta^2$. The classical
conserved densities are then $\beta^2$-independent and
the quantum corrections in the scheme $(m,\,\fh)$
are a polynomial in $\beta^2$. We shall be interested
in the 1 loop corrections to $K_n(\beta)$ characterizing
the eigenvalues (3.21). This amounts to calculating the
1 loop corrections to the Greens function
$G^{(n)}_a(p)$ appearing in (3.20). It is convenient to pick
the renormalization scheme $(m,M/m=\xi)$ in which the
fundamental tadpole (2.8) vanishes. In this case only the
parts $J^{(n,3)}+J^{(n,2)}$ of the conserved densities
(i.e. those of power 2 and 3 in $\d\Phi_a$) are needed for
the 1 loop corrections. Notice that to lowest order the
schemes $(m,\,\fh)$ and $m,\,M/m =\xi)$ coincide, so that
one can use the mCFT-form of the conserved densities as
input for the calculation. The Feynman diagrams contributing
to $G^{(n)}_a(p)$ are listed in Fig. 1. Diagrams of type
$A$ and $B$  arise from the classical part of
$J^{(n,3)}$ and $J^{(n,2)}$, respectively. Diagrams of type
$C$ involve contributions from the wave function
renormalization and from the first quantum corrections
of $J^{(n,2)}$. The $\otimes$ indicates the operator
insertion at zero momentum.
The wave function renormalization constants are known
exactly \cite{DDeV}. To $o(\beta^4)$ they read
\be
Z_a(\beta) =1-\frac{\beta^2(h-2)}{4\pi h} +
\frac{\beta^2(h-2a)}{4h^2}\cot\frac{a\pi}{h}+o(\beta^4)\;,
\ee
where $h=r+1$. Moreover from (4.1), (4.7) also
$J^{(n,2)}$ is known to $o(\beta^4)$. The contributions
from diagrams of type $C$ can therefore directly be
evaluated. For type $A$-diagrams one needs the explicit
form of $J^{(n,3)}$, which is not known for generic $n$.
\begin{center}
\mbox{
\setlength{\unitlength}{1mm}
\begin{picture}(140,80)
\thicklines
\unitlength1mm
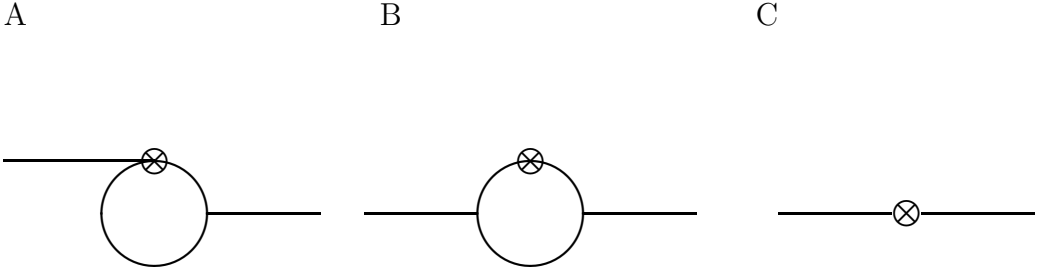
\put(-10,75){\makebox(0,0)[l]{Fig.1: Feynman diagrams for
 $o(\beta^2)$ contribution to $G^{(n)}_a(p)$}}
\put(-5,15){
\begin{picture}(50,60)
\put(5,45){A}
\put(25,20){\circle{20}}
\put(25,27){\line(-1,0){20}}
\put(32,20){\line(1,0){15}}
\put(25,27){\makebox(0,0){$\bigotimes$}}
\end{picture}
}
\put(45,15){
\begin{picture}(50,60)
\put(5,45){B}
\put(25,20){\circle{20}}
\put(18,20){\line(-1,0){15}}
\put(32,20){\line(1,0){15}}
\put(25,27){\makebox(0,0){$\bigotimes$}}
\end{picture}
}
\put(95,15){
\begin{picture}(50,60)
\put(5,45){C}
\put(23,20){\line(-1,0){15}}
\put(27,20){\line(1,0){15}}
\put(25,20){\makebox(0,0){$\bigotimes$}}
\end{picture}
}
\end{picture}
}
\end{center}
\vspace{-2cm}

\noindent
In the following we will consider $n=1,2$, where the
densities have been listed in (3.6). This is sufficient
for our purposes because in section 5 an exact expression
for $K_n(\beta)$ will be derived, for which the 1-loop PT
serves only as a check. In particular, $n=1$ is a good
check on the consistency of the set-up, because
$I^{(1)}(a)$ should reproduce the known first order
corrections to the quantum masses. The case $n=2$ then
provides a non-trivial check on the n-dependence.
The results obtained are consistent with the expression
\be
\In(a) =\In(a)_{class} \left(1-\frac{\beta^2}{8(r+1)}\,n\,
\cot\frac{a\pi n}{r+1} +o(\beta^4)\right)\;.
\ee
Consider first the case n=1. The eigenvalue should
reproduce the quantum masses via
\be
2\sqrt{2} I^{(1)}(a) =(m_a)_{phys}\;,
\ee
where $(m_a)_{phys}$ now are the physical masses
evaluated in the scheme $(m,\,M/m=\xi)$. To lowest order
(see the last ref. in \cite{AFZ})
\be
(m_a)_{phys}=m_a\left(1- \frac{\beta^2}{8(r+1)}
\cot\frac{a\pi}{r+1} +o(\beta^4)\right)\;.
\ee
On the other hand calculating $I^{(1)}(a)$ from (3.20),
(3.19) results in
\bas
2\sqrt{2} I^{(1)}(a) \is(m_a)_{phys}\,Z_a^{-1}\left[1+
\frac{1}{i p_-^2}\,B^{(1)}(a) \right]\;,\nonum
B^{(1)}(a)\is\sum_{bc}\lb_{abc}\lb^*_{abc}\,
Q_2(m_b,m_c|m_a)\;,
\eas
where $Q_2$ is defined and evaluated in appendix B.
Performing the sum one finds
\bas
&& \frac{1}{i p_-^2} B^{(1)}(a) =\frac{\beta^2}{4\pi h}
\left(2-h +\frac{4\pi(h-2a)}{h}\cot\frac{a\pi}{h}\right) +
o(\beta^4)\;,
\eas
which precisely cancels against the $o(\beta^2)$
contribution coming from $Z_a^{-1}$. Thus, to $o(\beta^2)$
one has indeed (4.10).

For $n=2$ we have verified the relation (4.10) for $r=2$
and $3$.%
\footnote{The $r=2$ case was also done in \cite{DGZ}.
The results do not coincide because in \cite{DGZ} the
eigenvalues were defined w.r.t. the bare mass scale.
In this case $I^{(1)}(a)$ does not
reproduce the quantum masses.}
We illustrate the calculation for $r=2$; the case $r=3$
is done similarly. From (3.6) the conserved density is
(after the rescaling $\Phi_a\ra \beta^{-1}\Phi_a$)
\be
J^{(2)} =\mp\frac{i}{3\sqrt{3}}\left(\mbox{$\Phi_1'$}^3-
\mbox{$\Phi_2'$}^3\right) \pm \frac{i}{2\sqrt{3}}
\left(1+\frac{\beta^2}{4\pi}\right)\left(\Phi_1''\Phi_2' -
\Phi_1'\Phi_2''\right)\;,
\ee
where $\Phi_a'=\d\Phi_a$ etc.. The 3-point vertex is
$V_3[\Phi]=\frac{3m^2}{\beta^2}(\Phi_1^3 +\Phi_2^3)$.
{}From (3.20) one obtains
\ba
I^{(2)}(1)\is\frac{1}{2\beta}\,Z_1^{-1}(m^2_1)_{phys}
\left[\frac{1}{2\sqrt{3}}
\left(1+\frac{\beta^2}{4\pi}\right)\right.\nonum
&& \left.\sspace\bspace -\frac{\beta^2}{(2\pi)^2i(p_-)^3}
\left(a\,A^{(2)}(1) +b\,B^{(2)}(1)\right)\right]\;,
\ea
where the momentum $p_-$ is on mass-shell
$p^2=(m_1^2)_{phys}$. The integrals
\bas
&& A^{(2)}(1) =m^2\int d^2 k
\frac{k_-(p_- -k_-)p_-}{(k^2-3m^2)[(k-p)^2 -3m^2]}\\
&& B^{(2)}(1) =m^4\int d^2 k
\frac{k_-^3}{(k^2-3m^2)^2[(k-p)^2 -3m^2]}
\eas
are $m$-independent, and $a,b$ are the corresponding
symmetry factors. From (4.13) and the form of $V_3[\Phi]$
one finds $a=0,\;b=3\sqrt{3}/2$ and  the evaluation of
(B.1) gives
\bas\nspace
&& B^{(2)}(1) =m^4 Q_3(3m^2,3m^2|3m^2)
=\frac{i\pi (p_-)^3}{27}
\left[4-\frac{7\pi}{3\sqrt{3}}\right]\;.
\eas
Inserting into (4.14) one obtains
\be
I^{(2)}(1) =\left(I^{(2)}(1)\right)_{class}
\left(1+\frac{\beta^2}{12\sqrt{3}}\right)=
-I^{(2)}(2)\;,
\ee
consistent with (4.10).
\pagebreak
\newsection{Vertex operator construction of $K_n(\beta)$}
\vspace{-15mm}

\newsubsection{Realizations of the ZF algebra}
In order to obtain an exact expression for $K_n(\beta)$
we use the bootstrap S-matrix as an additional input.
Since the scattering operator commutes with
all the conserved charges $[S\,,\,\In]=0$, one expects
the $S$-matrix also to carry information about their
spectrum. Indeed, the intertwining
concept is just that of a Zamolodchikov-Faddeev (ZF)
operator. If $\Zt{a}$ denotes the ZF operator creating
an asymptotic 1-particle state $|a(\theta)\!\ket$ of type
$a$ and rapidity $\theta$, one has by definition
\setcounter{abc}{1}
\renewcommand{\theequation}%
{\thesection.\arabic{equation}\alph{abc}}
\ba
&& \ZZ{a}{a}\,\ZZ{b}{b} = S_{ab}(\theta_a -\theta_b)\,
\ZZ{b}{b}\,\ZZ{a}{a}\\
\addtocounter{equation}{-1}
\addtocounter{abc}{1}
&& [\In\,,\,\Zt{a}]=e^{-n\theta}\,\In(a)\,\Zt{a}\;,\\
\addtocounter{equation}{-1}
\addtocounter{abc}{1}
&& \left[ K\,,\; \Zt{a}\right] =
\frac{d}{d\theta}\,\Zt{a} \;,
\ea
\renewcommand{\theequation}{\thesection.\arabic{equation}}
$\!$where $\In(a)$ is the eigenvalue of $\In$ on
$|a(\theta)\!\ket =\Zt{a}|\Omega\!\ket$ and $K$ is the
generator of Lorentz boosts. We will call an associative
algebra with generators $\Zt{a},\;\theta\in\C$,
$\In,\;n\in E$ and $K$ subject to the relations (5.1)
(and possibly others) a {\em ZF algebra} $Z(S)$ associated
with $S$. Usually, a ZF algebra is supposed to act on the
space of scattering states of the theory $\Sigma_{in/out}$.
These are Fock spaces but the relation to the fundamental
fields of the theory is elusive in general. It is therefore
more useful to construct realizations
\bas
&& \rho :Z(S)\rra \pi\;,
\sspace\rho(A\,B)=\rho(A)\,\rho(B)\;,
\eas
of (5.1) on
some auxilary Fock space $\pi$, on which the $\Zt{a}$'s
act as generalized vertex operators. This, of course, can
be done in many ways and one will choose the realization
according to purpose. The following two realizations are of
particular interest.
\begin{itemize}
\item[{(1)}] The realization $\rho_I$ adapted to the
conserved charges. In this case $\pi =\C[x_n,\;n\in E]$
and the realization is defined by
\be
\rho_I(\In) =\ddx{n},\;n\in E\;.
\ee
We will show below that, supplemented by the invariance
under some involution $\omega$, the
the relation (5.2) fixes the realization
almost uniquely. In particular, the 2-point function
coincides with the phase of the minimal form factor (2.27)
\be
\langle 0|\rho_I(Z_a)(\theta_1)\,
\rho_I(Z_b)(\theta_2)|0\rangle
= e^{i Im\,f_{ab}(\theta_1-\theta_2)}.
\ee
Once the realization is known, the
eigenvalues $\In(a)$ can be obtained from the relation
(5.1.b).
\item[{(2)}] The realization $\rho_F$ adapted to the
form factor equations. In this case one has some freedom
in the choice of the Fock space, but usually again the
Fock space of a single free boson will be sufficient i.e.
$\pi =\C[y_n,\;n\in E]$. The main requirement now is that
the two point function $G_{ab}(\theta_1-\theta_2) :=
\langle 0|\rho_F(Z_a)(\theta_1)\,
\rho_F(Z_b)(\theta_2)|0\rangle$
satisfies the conditions%
\footnote{It is worth emphasizing that $G_{ab}(\theta)$
is {\em not} supposed to satisfy $G_{ab}(\theta +2\pi i)=
e^{2\pi i l}G_{ba}(-\theta)$ for some $l \in \R$.}
\begin{itemize}
\item $G_{ab}(\theta)$ is analytic for
$\mbox{Im}\,\theta \leq 0$, except for a simple pole at
$\theta =-i\pi$.
\item $G_{ab}(\theta)$ is bounded in the lower half
plane, $G_{ab}(\theta)= O(1),\;\theta\ra\infty,\;
\mbox{Im}\,\theta\leq 0$.
\end{itemize}
Again this fixes the realization essentially uniquely.
Suppose now the realization to be given and let $\Lambda$
be any linear operator on $\pi$ satisfying
\ba
&& \Lambda \,\Zt{a} =e^{2\pi i l}\Zt{a}\, \Lambda\nonum
&& e^{\theta K}\,\Lambda\, e^{-\theta K}=
e^{\theta s}\,\Lambda\;,
\ea
for some $l\in \R,\;s\in \Z$. According to
\cite{Luk}, eqn. (3.9) the function
\be
F^{\Lambda}_{a_n\ldots a_1}(\theta_1,\ldots,\theta_n) =
\Tr_{\pi}\left[e^{2\pi i \rho_F(K)}\,\Lambda \,\rho_F
\left(Z_{a_n}(\theta_n)\ldots Z_{a_1}(\theta_1)\right)\right]
\ee
then formally satisfies the form factor equations.
\end{itemize}
Both realizations are designed to solve some aspect of
the full problem (non-perturbative construction of the
QFT) but fail to incorporate others. The construction
(2) does not specify which of the solutions (5.4)
correspond to the form factors of some given local operator
$\cO$. If one interprets $l$ and $s$ in (5.4) as the
locality index (relative to the fundamental field) and
the spin of $\cO$, respectively, any solution of (5.4)
can apparently be taken to represent $\rho_F(\cO)$.
Clearly some additional dynamical input is needed to
determine the realizations $\rho_F(\cO)$ of local operators.

Conversely, in the realization $\rho_I$ one fixes the image
of at least an infinite subset of local operators, namely the
conserved charges, via (5.2). This essentially fixes the
realization and allows one to determine the eigenvalues
of the conserved charges. On the other hand the two point
function (5.3)
will in general fail to have simple analyticity properties
in $\theta_1-\theta_2$ and the operators $\rho_I(Z_a)$ will
not be useful to obtain solutions of the form factor
equations. We expect that important technical progress can
be made by understanding the relation between the
realizations (1) and (2) (and possibly others). In
particular, one should find $\rho_F(\In)$.

\newsubsection{Construction of $\rho_I$}
Recall that the quantization has been done w.r.t. the
$\dmi$-lightcone dynamics. In particular the conserved
charges $\In$ and their proposed realization $\rho_I$
refer to the choice of the $\dmi$-dynamics. Of course
one could also have chosen the $\dpl$-lightcone
dynamics and the result for the eigenvalues should be
the same. In other words, if we define $\omega$ to be
the involution which maps the theory quantized w.r.t.
the $\dmi$-dynamics onto that quantized w.r.t. the
$\dpl$-dynamics, one should have
\be
\omega\left(\In(a)\right)=\In(a)\;.
\ee
On the imaginary axis in rapidity space the action of
$\omega$ should correspond to $e^{-i\theta}\ra
c^2 e^{i\theta}$, where the real constant $c$ is related to
the overall mass scale. For the rapidity factors
$e^{-in\theta}$ accompanying $\In(a)$ one sees that
changing the sign of $\theta$ formally amounts to the same
as changing the sign of $n$. We can thus introduce an
infinite dimensional Heisenberg algebra
\be
[a_{-m},a_n]=i\,k\,\delta_{m,n}\;,\sspace m,n\in E\;,
\ee
and associate the positive modes with the charges $\In$
and the negative modes with $\omega \In$. Consistency
then requires
\ba
\omega(a_m)= -c^{2m} a_{-m}\;,\sspace &&
\omega \left([a_{-m},a_n]\right)
 = [\omega(a_{-m}),\omega(a_n)] \nonum
\omega(a_{-m})=-c^{-2m}a_m\;,\sspace &&
\omega(z) =z^*\;,\;\;z\in \C\;.
\ea
i.e. $\omega$ is a linear anti-involution. Let $\pi$ denote
the Fock space $\C[a_{-n},\;n\in E]$ with vacuum $|0\ket$.
On $\pi$ there exists a unique bilinear form
$\bra\;,\;\ket$ s.t. $\omega$ is contravariant w.r.t. it, i.e.
$\bra A,B\ket = \bra 0,\omega(A) B\ket$.
In particular $\bra a_{-1},a_{-1}\ket=-i\,c^{-2}\,k$.
The defining relation (5.1.b) then becomes%
\footnote{To simplify the notation we will write $\Zt{a}$
instead of $\rho_I(\Zt{a})$ when no confusion is possible.
For real $\theta$ these are unbounded operators on $\pi$.
The following formulae can be justified e.g. through
analytic continuation from complex values of $\theta$.}
\be
[a_n\,,\,\Zt{a}]=e^{-in\theta}\,\In(a)\,\Zt{a}\;,
\sspace n\in E\;.
\ee
If we require that the realization $\rho_I$ is $\omega$-%
invariant,
\be
\omega\left(\Zt{a}\right) =\Zt{a}\;,
\ee
this implies that
\be
[a_{-n}\,,\,\Zt{a}]= -c^{2n}\,e^{in\theta}\,\In(a)\,\Zt{a}\;,
\sspace n\in E\;.
\ee
Slighty extending a well-known Lemma (\cite{Kac},
Lemma 14.5), we infer that a linear operator
on (the formal completion of) $\pi$ satisfying the
commutation relations (5.9), (5.10) is a generalized vertex
operator i.e. of the form
\ba
&& \rho_I(Z_a)(\theta) = :e^{\Upsilon_a(\theta)}:\;,\nonum
&& \Upsilon_a(\theta) =
\sum_{n\in \pm E}d_n(a)\,a_n\, e^{in\theta}\;,\sspace
a=1,\ldots ,r
\ea
where $:\;\;:$ denotes normal ordering w.r.t. the
Heisenberg algebra (5.7). The coefficients $d_n(a)$ are to
be determined s.t. the equations (5.1.a), (5.10) holds.
{}From $\omega\Upsilon_a(\theta)=\Upsilon_a(\theta)$ one
finds
\be
d_{-n}^*(a) =-d_n(a)\,c^{2n}\;,\sspace n\in E\;.
\ee
To implement the relation (5.1.a) first notice that the
compatibility of (5.1) with the S-matrix bootstrap
equations (2.17) imposes the consistency consitions
\ba
&&Z_{\bar{a}}(i\pi -\theta)
= c_a\,Z_a(-\theta)^{-1}\nonumber\\
&& Z_a(\theta +i\eta(a))\,
Z_b(\theta +i\eta(b))\,
Z_c(\theta +i\eta(c)) =c_{abc}\;,
\ea
where $c_a$ and $c_{abc}$ are constants. We may assume these
constants to equal 1. In terms of the fields
$\Upsilon_a(\theta)$ the conditions (5.14) then become
\bas
&& \Upsilon_a(i\pi -\theta) =
\Upsilon_{\bar{a}}(-\theta)\;,\nonumber\\
&& \sum_{l=a,b,c}\, \Upsilon_l(\theta + i\eta(l)) =0\;,
\eas
Using the results of appendix A, these equations enforce
$d_n(a) =d_n q_a^{(n)}\;,n\in\pm E$ for some $a$-independent
constants $d_n$. Consider now
\be
Z_a(\theta_1)\,Z_b(\theta_2) = \,
:\,Z_a(\theta_1)\,Z_b(\theta_2)\,:
\exp\left( i\,k\sum_{n\in E}d_n(a) d_{-n}(b)\,
e^{in(\theta_1-\theta_2)}\right)\;.
\ee
The monodromy of $Z_a(\theta_1)\,Z_b(\theta_2)$ will
therefore be given by the meromorphic continuation of the
function $\exp\left( 2i\,k\sum_{n\in E}d_n(a) d_{-n}(b)\,
e^{in(\theta_1-\theta_2)}\right)$. Comparing with the
expansion (2.31) of the scattering phase one reads off the
condition
\be
\frac{1}{n}\,D_n\, q_a^{(n)} q_b^{(n)} =
2 k \,d_n(a)\, d_{-n}(b) =
2 k\, d_n d_{-n}\, q_a^{(n)} q_b^{(-n)}\;,\;\;n\in E\;.
\ee
Since $q_b^{(-n)} =-q_b^{(n)}$ this implies
$D_n =-2n k\,d_n d_{-n}=-D_{-n}$, so that by (5.13)
\be
d_n = i\epsilon_n\,c^{-n}\, \sqrt{\frac{D_n}{2n k}}\;,
\ee
where $\epsilon_n$ is a sign to be determined later.
In summary we have obtained a 1-parameter family of
realizations of (5.1) from the conditions (5.2), (5.10)
\ba
&& \rho_I(Z_a)(\theta) = :e^{\Upsilon_a(\theta)}:\;,\nonum
&& \Upsilon_a(\theta) = i \sum_{n\in \pm E}
\epsilon_n\, c^{-n}\, \sqrt{\frac{D_n}{2n k}}\;
q_a^{(n)}\,a_n\, e^{in\theta}\;,\sspace
a=1,\ldots ,r\;.
\ea
Combining (5.15) and (5.17) one sees that this realization
has the feature (5.3). The Lorentz boost operator is
realized as
\be
\rho_I(K) =-\frac{1}{k}\sum_{n\in E} n a_{-n} a_n\;
\ee
and is $\omega$-invariant. Having constructed the
realization we can now use the the defining relation
(5.1.b) -- or equivalently (5.9), (5.11) -- to find the
eigenvalues $\In(a)$. Equating the expressions obtained
from (5.9) and
(5.11) fixes the sign $\epsilon_n =(-)^{sign(n)+1}
=-\epsilon_{-n}$.(i.e. $q_a^{(n)}$ in (5.18) gets replaced
by $q_a^{(|n|)}$.) Comparison with the generic form (3.21)
then gives the condition
\be
\In(a) = c^n \sqrt{\frac{k\,D_n}{2n}}\; q_a^{(n)}
\stackrel{\mbox{!}}{=} K_n(\beta)
\left(\frac{m^2 e^T}{2\beta^2}\right)^{n/2}\; q_a^{(n)}\;.
\ee
This fixes
\be
c= \left(\frac{m^2 e^T}{2\beta^2}\right)^{1/2}\;,\sspace
K_n(\beta) = \sqrt{\frac{k\,D_n}{2n}}\;.
\ee
Further, since $D_n$ is known, the unknown function $K_n(\beta)$
has been determined up to the constant $k$, appearing in the
commutation relations (5.7). Expanding $D_n$ and matching
against the perturbative result (4.10) one sees that $k$
actually has to be a function of $\beta$ of the form
\be
k(\beta) = h\left(
1+\frac{\beta^2}{4\pi} +o(\beta^4)\right)\;.
\ee
Since $k(\beta)$ is $n$-independent, the higher order
terms in (5.22) can be found by comparison with the exact
quantum masses of Destri and DeVega\cite{DDeV}. When
written in a RG invariant form, these are
\begin{samepage}
\be
(m_a)_{phys}= 2\sqrt{2} I^{(1)}(a)=
\left(\frac{m^2\,e^T\,h D_1}{\pi B}\right)^{1/2}\,
q_a^{(1)}\;,
\ee
so that $k(\beta)=h\beta^2/2\pi B$, i.e.
(5.22) is in fact exact.
\end{samepage}

In summary we have obtained the following result: There
exists a 1-parameter family of $\omega$-invariant
realizations of the ZF algebra s.t. $\rho_I(\In)=\ddx{n} =
a_n$. The free parameter $c$ corresponds to
the ambiguity in setting the physical mass scale through
a (renormalization group invariant) combination of the
bare mass and the normal ordering mass and is given
explicitely in (5.21). The parameter $c$ also sets the
scale for the involution $\omega(a_n) =-c^{2n} a_{-n}$
and the commutation relations are $[a_{-m},a_n] =
ih(1+\beta^2/4\pi) \delta_{m,n}$. On the (formal completion
of the) corresponding Fock space representation the
asymptotic multi-particle states are realized as
generalized coherent states
\bas
\rho_I\left(|a_N(\theta_N),\ldots, a_1(\theta)\ket \right)
=\rho_I(Z_{a_N})(\theta_N)\ldots
\rho_I(Z_{a_1})(\theta_1)|0\ket \;.
\eas
These are simultaneous eigenstates of the conserved charges
$\In,\;n\in E$ and their eigenvalues decompose into a sum of
eigenvalues $\In(a)$ on the single particle states.
For these eigenvalues we have the exact result
\be
\In(a) =
\left(\frac{ m^2 e^T}{2\beta^2}\right)^{n/2} \beta
\left[\frac{h \,\sin\frac{\pi n}{2h}B
                           \sin\frac{\pi n}{2h}(2-B)
                      }{4\pi n\,B\,\sin \frac{\pi n}{h}}
           \right]^{1/2}\; q_a^{(n)}\;,\sspace
n\in E \;.
\ee
Here $h$ is the Coxeter number, $B$ is the effective
coupling (2.22) and $q^{(n)}_a$ are the eigenvectors of
the Cartan matrix. The tadpole function $T$ enters through
the RG invariant combination $m^2e^T$.
\newsection{Conclusions}
The equation (5.24) solves the diagonalization problem of the
conserved charges $\In$ in a QFT context, using the
bootstrap S-matrix as an input. Conversely, combining
equation (5.24) with (2.31) the scattering phase can also be
re-expressed in terms of the eigenvalues of the conserved
charges via
\be
\delta_{ab}(\theta) = \frac{4\pi B}{h\beta^2}\sum_{n\in E}
c^{-2n} \In(a)\,\In(b)\;
 e^{n\theta}\;,\bspace\mbox{Re}\,\theta<0\;.
\ee
This means that an independent justification of the
eigenvalues (5.24) would also provide a dynamical justification
of the conjectured bootstrap S-matrix. Since UV finite
expressions for the conserved charges can be constructed
directly from CFT techniques, this amounts to addressing the
diagonalization problem on suitable representation spaces
of the associated chiral algebra. For the real coupling
models the Fock space of $\mbox{rank}(g)$ free bosons
should be appropriate.

Generally one is lead to consider the diagonalization
problem of the conserved charges on the Verma modules of
$W(g)$. This amounts to the calculation of generalized
Kac determinants in the following sense. Let $V(\lb)$
be a Verma module of $W(g)$, where $\lb$ are the parameters
of the highest weight state. Let $V_N$ denote the
subspace of degree $N$ w.r.t. the $L_0$-grading
($L_0$ being the zero mode of the Virasoro subalgebra)
with basis $v_{\alpha},\;1\leq\alpha \leq \dim V_N=:K$.
If $\bra\;,\;\ket$ denotes the usual contravariant
hermitian form on $V_N$, the Kac determinant formula
(see \cite{BS} and references therein) gives an
expression for
\be
\det\left(\bra v_{\alpha},
\,L_0\, v_{\beta}\ket\right)_{1\leq \alpha,\beta\leq K}\;
\ee
as a function of $\lb$ and the central charge.
The insertion of $L_0$ of course just gives rise to
an overall factor and is usually omitted. From the
viewpoint of $W$-algebras however the conserved charges
$\In$ arise as an infinite dimensional abelian subalgebra
of $W(g)$. Moreover one can construct a basis of $W(g)$ for
which $I^{(1)}=L_0$ (rather than $I^{(1)}=L_{-1}$) so that
the charges $\In$ preserve the $L_0$ graduation of the
Verma modules $[L_0,\In]=0,\;n\in E$\cite{Carg}.
{}From this viewpoint, the choice of the lowest conserved
charge $I^{(1)}=L_0$ in (6.2) is not preferred and one can
study `generalized Kac determinants' of the form
\be
\det\left(\bra v_{\alpha},
\,\In\, v_{\beta}\ket\right)_{1\leq \alpha,\beta\leq K}\;,
\sspace n\in E\;.
\ee
Finding an explicit expression for (6.3) would solve the
diagonalization problem of the conserved charges on
Verma modules.

For the ATs one should study similar determinants on the
Hilbert space of the CFT characterizing its UV behaviour.
Let $v_{\alpha,\bar{\alpha}},\;1\leq \alpha\leq K,\;
1\leq \bar{\alpha}\leq \overline{K}$  be a basis of the
subspace of degree $N$ in the CFT (w.r.t. the $L_0 +
\overline{L}_0$ graduation) and consider
\be
\det\left(\bra v_{\alpha\bar{\alpha}},
\,\In\, v_{\beta\bar{\beta}}\ket
\right)_{ {1\leq \alpha,\beta\leq K        \atop
          1\leq \bar{\alpha},\bar{\beta}\leq\overline{K} }
            }\;,
\sspace n\in E\;.
\ee
\begin{samepage}
On general grounds we expect that the large $N$ asymptotics
of these determinants is governed by factors whose zeros
are proportional to sums of factors $K^n\,\In(a)$ for some
constant $K$.
\vspace{1cm}

\noindent{\tt Acknowledgements:} I wish to thank P.~Weisz
for some valuable discussions.
\end{samepage}
\vspace{2cm}

\noindent{\large\bf  Appendix}
\setcounter{section}{0}
\vspace{-1cm}

\newappendix{Orbits of the bicolored Coxeter element}
Let $g$ be a simple simply laced%
\footnote{The following
results hold with minor modifications also for non simply
laced algebras. The status of the associated bootstrap S-matrices
is however not clear yet.}
Lie algebra.
The simple root system of $g$ admits a bicoloring
s.t. roots of different colors are orthogonal.
Let $\Omega_+$ denote the product of the Weyl
reflections in the `white' simple roots $\alpha_a$
endowed with the color value $c(a)=+1$ and
$\Omega_-$ the product of the Weyl reflections
in the `black' simple roots
endowed with the color value $c(a)=-1$.
Then $\Omega =\Omega_-\Omega_+$ is a preferred
Coxeter element of $g$. In particular, its
eigenvectors are labeled by the exponents
$\{s_1,\ldots ,s_r\}$ of $g$
\be
\Omega e^{(s)} = e^{\frac{2\pi i s}{h} } e^{(s)}\;,
\bspace e^{(s)}\cdot e^{(t)}= \delta_{s+t, h}\;,
\ee
where $s \in \{s_1,\ldots, s_r\}$ and $h$ is the
Coxeter number. The components of some real
$\lambda \in h^{\ast}$ are given by
\ba
&& \lambda =\sum_s \lambda^{(h-s)} e^{(s)},\;\nonumber\\
&&\lambda^{(s)} = (e^{(s)},\lambda),\sspace \;
\lambda^{(h-s)}=(\lambda^{(s)})^*\;.
\ea
It suffices to know the components of the simple roots
and the fundamental weights. The result is
(see e.g. \cite{Dor})
\ba
&& (e^{(s)},\,\lambda_a) =
\frac{i}{\sqrt{2}\sin\theta_s}
e^{-i\theta_s \left(\frac{1+c(a)}{2} \right) }
q_a^{(s)}\;,\nonumber\\
&& (e^{(s)},\,\alpha_a) =
-c(a)\sqrt{2}\,
e^{i\theta_s \left(\frac{1-c(a)}{2} \right) }
q_a^{(s)}\;,
\ea
where $\theta_s = \pi s/h$ and $q_a^{(s)}=c(a)q_a^{(h-s)}$
is the normalized eigenvector of the Cartan matrix
defined by
\be
\sum_a a_{ab}\, q_{b}^{(s)} = 2(1-\cos \theta_s)\, q_a^{(s)}\;,
\sspace q^{(s)}\cdot q^{(t)} =\delta_{s,t}\;.
\ee
This allows to evaluate inner products of the form
\be
\left(\lambda, \Omega^{-p}\mu\right) =
\sum_s \omega^{ps} \lambda^{(h-s)}\mu^{(s)}
\ee
 In particular
\be
\left(\lambda,\Omega^{-p}\mu\right)
= \sum_s q_a^{(s)}\,q_b^{(s)}\;
e^{i\theta_s(2p+c(a,b))} d_s
= \sum_s q_a^{(s)}\,q_b^{(s)}\;\tilde{d}_s\;,
\ee
\be \!\!\!\!\!\!\!\!\!\!\!\!\!\nspace
\begin{array}{lll}
\sspace d_s =
\frac{{\dis 1}}{{\dis 2\sin^2\theta_s} }\;;&\sspace
\tilde{d}_s=\frac{{\dis \cos(2p +c(a,b))\theta_s}}
                 {{\dis 2\sin^2\theta_s } }\;;
&\sspace \mbox{for}\;\;\lambda,\;\mu =\lambda_a,\;\lambda_b\;
\nonumber\\
\sspace d_s =\frac{{\dis ie^{i\theta_s} } }
{{\dis \sin\theta_s}} \;;&\sspace
\tilde{d}_s=\frac{{\dis \sin(2p+1 +c(a,b))\theta_s} }
                 {{\dis \sin\theta_s} }\;;
&\sspace \mbox{for}\;\;\lambda,\;\mu =\lambda_a,\;\gamma_b\;
\\
\sspace d_s =2\;;&\sspace
\tilde{d}_s=2\cos(2p +c(a,b))\theta_s\;;
&\sspace \mbox{for}\;\;\lambda,\;\mu =\gamma_a,\;\gamma_b\;,
\end{array}
\ee
where $\gamma_a =c(a)\alpha_a,\;c(a,b)=(c(a)-c(b))/2$
and the symmetric form of (B7) displays that the r.h.s.
of (B6) is real. Set
\be
A_{\lambda,\mu}(\theta) =\prod_{p=1}^h
\left( 1- e^{\frac{i\pi}{h}(2p-c(a,b))}
    e^{\theta}\right)^{(\lambda,\Omega^{-p}\mu)}\;.
\ee
Then
\be
\ln\,A_{\lambda\mu}(\theta) = -\sum_{n\in E}
\frac{h}{n}\, d_n\, q_a^{(n)}\,q_b^{(n)}\;e^{n\theta}\;,
\ee
with the subcases (B7).

\noindent {\em Remark:} $A_{\gamma_a,\gamma_b}$ is the
2-soliton interaction constant. For $g=A_r$ it reduces to
\be
A_{\gamma_a\gamma_b} =
\frac{ \sinh\left( \frac{\theta}{2} +
                  \frac{i\pi(a-b)}{2(r+1)}\right)
       \sinh\left( \frac{\theta}{2} -
                  \frac{i\pi(a-b)}{2(r+1)}\right)
     }
     { \sinh\left( \frac{\theta}{2} +
                  \frac{i\pi(a+b)}{2(r+1)}\right)
       \sinh\left( \frac{\theta}{2} -
                  \frac{i\pi(a+b)}{2(r+1)}\right)
     }
\ee
The quantity $A_{\gamma_a\lambda_b}$ enters the bootstrap
S-matrices and $A_{\lambda_a\lambda_b}$ has been used
in \cite{CorDor} for the construction of vertex operators.

Let $\Omega=\Omega_-\Omega_+ =r_{i_r}\ldots r_{i_1}$
denote a reduced expression for the bicolored
Coxeter element. Set
\bas
\nspace && \Delta_+^{\Omega} =\{\alpha \in \Delta_+|
\Omega(\alpha) <0\}\;,
\eas
which by $|\Delta_+^{\Omega}|=l(\Omega)=r$ has r elements.
An explicit enumeration is
\ba
&\Delta_+^{\Omega}& =\left\{\alpha_{i_r},\;
r_{i_r}\alpha_{i_r-1},\ldots ,r_{i_r}...r_{i_2}\alpha_{i_1}
\right\} \nonumber\\
&                 & =\left\{ (1-\Omega^{-1})\lambda_a\;,\;\;
1\leq a \leq r\right\}\;.
\ea
Consistent with $|\Delta|=2|\Delta_+|= 2\frac{1}{2} r h =
\mbox{dim}\,g -r$ the root system decomposes
into $r$ disjoint orbits of $\Omega$ each of
which is $h$-dimensional. Let $\Omega_a =
Z_h\cdot (1-\Omega^{-1})\lambda_a$ denote the
orbit of the $a$-th element in (B11) under the
action of the cyclic group $Z_h$ generated by
$\Omega$. Then $\Delta=\Omega_1 \oplus \ldots
\oplus \Omega_r$. The orbits $\Omega_a$ have the
following properties:
\begin{itemize}
\item[$i.$]
$\gamma_a =c(a)\alpha_a \in \Omega_a \;$.
\item[$ii.$]
If $\Omega_a$ is an orbit, so is $-\Omega_a$ and
hence has to coincide with some $\Omega_{\bar{a}}$,
where `$\,\bar{}\,$' denotes an idempotent permutation
of $\{1,\ldots,r\}$. Explicitely \cite{Olive}
\bas
\gamma_{\bar{a}} = -\Omega^{-\frac{h}{2} +
 \frac{c(a)-c(\bar{a})}{4}}
\gamma_{a} \;, \bspace c(\bar{a})c(a) =(-)^h\;.
\eas
\item[$iii.$] For $(a,b,c)\in \{1,\ldots, r\}^3$ the
equivalent equations
\ba
&&\sum_{l=a,b,c} \Omega^{\,\zeta(l)}\,\gamma_l =0 \nonum
&&\sum_{l=a,b,c} \Omega^{-\zeta(l)}\,\lambda_l =0
\ea
are called `fusing equations' for the process
$a,b\rightarrow \bar{c}$. The triplet
$(\zeta(a),\,\zeta(b),\,\zeta(c)) \in \Z^3/\Z$
is called a solution. For given $(a,b,c)\in
\{1,\ldots ,r\}^3$ (B12) has either none or two
independent solutions. If
$(\zeta(a),\,\zeta(b),\,\zeta(c))$  is one solution
then \newline $(\zeta'(a),\,\zeta'(b),\,\zeta'(c))$,
$\zeta'(l)=-\zeta(l) +(c(l)-1)/2$ is the second\cite{Olive}.
The projections of (B12) onto the eigenstates of $\Omega$
are given by
\be
\sum_{l=a,b,c} e^{\pm is\eta(l)} q_l^{(s)} =0\;,
\ee
respectively, where $\eta(l) = -\frac{\pi}{h}(2\zeta(l) +
\frac{1-c(l)}{2})$. Since $\eta'(l)=
-\frac{\pi}{h}(2\zeta'(l) +  \frac{1-c(l)}{2})=
-\eta(l)$ the two eqn.s (B13) correspond to the
two solutions of the fusing equations.
Charge conjugation corresponds to
$\eta(\bar{a})=\eta(a) -\pi$.
\item[$iv.$] Let $T_a$ denote a basis
of the Cartan subalgebra satisfying
$T_a^{\dagger} =T_{\bar{a}}\;,\mbox{Tr}(T_aT_b)=
\delta_{a\bar{b}}$ and let $E=\sum_{i=1}^r e_i +e_{-\theta}\;,
F=\sum_{i1}^r \check{n}_i f_i + e_{\theta}$
(with $\theta$ the highest root and $\check{n}_i$ the
dual Kac lables) denote the standard
regular elements of $g$. The 3-point coupling
$C_{abc} =\mbox{Tr}\left(\left[ [T_a,\,E]\,,\,
[T_b,\,F]\right],\; T_c\right)$ is non-zero iff
the fusing equation for $(a,b,c)$ has a nontrivial
solution\cite{Dor,Olive}.
\end{itemize}
\newappendix{Some Integrals}
Here we spell out some details needed for the perturbation
theory calculation in section 4.2. The required integrals
are
\bas
Q_n(a^2,b^2|p^2) \is\int d^2 k\frac{(k^-)^n}%
{[k^2-a^2 +i\epsilon]^2[(k-p)^2 -b^2 +i\epsilon]}\\
\is\int dk^+dk^-\frac{(k^-)^n}%
{[2k^+k^- -a^2 +i\epsilon]^2
[2(k^+-p^+)(k^-+p^-) -b^2 +i\epsilon] }\;.
\eas
Closing the contour in the lower half $k^+$ plane gives
\be
Q_n(a^2,b^2|p^2) =-i\pi\left[K_{n+1}(0)-K_{n+1}(p^-)-
p^-K_n(0)+p^-K_n(p^-)\right]\;,
\ee
where
\bas \nspace
&& K_n =\int_{-\infty}^{\infty}dk\frac{k^n}%
{2p^+k^2 +(b^2-a^2-p^2)k +a^2 p^-}\;.
\eas
In particular
\ba
&& Q_2(a^2,b^2|p^2) =\frac{-i\pi(p^-)^2}{p^4\Delta}
\left[p^2(a^2-b^2) +p^4 + \right. \nonum
&&\bspace +\left. \left((b^2-a^2-p^2)\frac{\Delta}{2}
 +p^2 a^2(p^2+b^2-a^2)\right) Q\right]-
\frac{i\pi(p^-)^2}{2p^4}\ln\frac{a^2}{b^2}\;,
\nonum
&& \Delta =-p^4+2p^2(a^2+b^2)-(a^2-b^2)^2\;,
\nonum
&& Q=-\frac{2}{\sqrt{\Delta}}\left[
\arctan\left(\frac{-p^2+b^2-a^2}{\sqrt{\Delta}}\right)-
\arctan\left(\frac{p^2+b^2-a^2}{\sqrt{\Delta}}\right)
\right]\;.
\ea

\end{document}